\documentclass[english,useAMS,usenatbib, onecolumn]{mn2e} 
\usepackage{lmodern}

\usepackage[T1]{fontenc}
\usepackage[utf8]{inputenc}
\usepackage{color}
\usepackage[usenames,dvipsnames]{xcolor}
\usepackage{amsmath}
\usepackage{amssymb}
\usepackage{graphicx}
\usepackage{esint}
\usepackage{placeins}
\usepackage{bigints}
\usepackage[authoryear]{natbib}

\usepackage{hyperref}

\definecolor{darkgreen}{rgb}{0.1,0.6,0.6}
\hypersetup{colorlinks = true,urlcolor=blue,citecolor=Blue}

\makeatletter
\usepackage{lmodern}\usepackage{epstopdf}

\let\jnl@style=\rm
\def\ref@jnl#1{{\jnl@style#1}}

\def\aj{\ref@jnl{AJ}}                   
\def\actaa{\ref@jnl{Acta Astron.}}      
\def\araa{\ref@jnl{ARA\&A}}             
\def\apj{\ref@jnl{ApJ}}                 
\def\apjl{\ref@jnl{ApJ}}                
\def\apjs{\ref@jnl{ApJS}}               
\def\ao{\ref@jnl{Appl.~Opt.}}           
\def\apss{\ref@jnl{Ap\&SS}}             
\def\aap{\ref@jnl{A\&A}}                
\def\aapr{\ref@jnl{A\&A~Rev.}}          
\def\aaps{\ref@jnl{A\&AS}}              
\def\azh{\ref@jnl{AZh}}                 
\def\baas{\ref@jnl{BAAS}}               
\def\bac{\ref@jnl{Bull. astr. Inst. Czechosl.}}
\def\caa{\ref@jnl{Chinese Astron. Astrophys.}}
\def\cjaa{\ref@jnl{Chinese J. Astron. Astrophys.}}
\def\icarus{\ref@jnl{Icarus}}           
\def\jcap{\ref@jnl{J. Cosmology Astropart. Phys.}}
\def\jrasc{\ref@jnl{JRASC}}             
\def\memras{\ref@jnl{MmRAS}}            
\def\mnras{\ref@jnl{MNRAS}}             
\def\na{\ref@jnl{New A}}                
\def\nar{\ref@jnl{New A Rev.}}          
\def\pra{\ref@jnl{Phys.~Rev.~A}}        
\def\prb{\ref@jnl{Phys.~Rev.~B}}        
\def\prc{\ref@jnl{Phys.~Rev.~C}}        
\def\prd{\ref@jnl{Phys.~Rev.~D}}        
\def\pre{\ref@jnl{Phys.~Rev.~E}}        
\def\prl{\ref@jnl{Phys.~Rev.~Lett.}}    
\def\pasa{\ref@jnl{PASA}}               
\def\pasp{\ref@jnl{PASP}}               
\def\pasj{\ref@jnl{PASJ}}               
\def\rmxaa{\ref@jnl{Rev. Mexicana Astron. Astrofis.}}%
\def\qjras{\ref@jnl{QJRAS}}             
\def\skytel{\ref@jnl{S\&T}}             
\def\solphys{\ref@jnl{Sol.~Phys.}}      
\def\sovast{\ref@jnl{Soviet~Ast.}}      
\def\ssr{\ref@jnl{Space~Sci.~Rev.}}     
\def\zap{\ref@jnl{ZAp}}                 
\def\nat{\ref@jnl{Nature}}              
\def\iaucirc{\ref@jnl{IAU~Circ.}}       
\def\aplett{\ref@jnl{Astrophys.~Lett.}} 
\def\apspr{\ref@jnl{Astrophys.~Space~Phys.~Res.}}
\def\bain{\ref@jnl{Bull.~Astron.~Inst.~Netherlands}}
\def\fcp{\ref@jnl{Fund.~Cosmic~Phys.}}  
\def\gca{\ref@jnl{Geochim.~Cosmochim.~Acta}}   
\def\grl{\ref@jnl{Geophys.~Res.~Lett.}} 
\def\jcp{\ref@jnl{J.~Chem.~Phys.}}      
\def\jgr{\ref@jnl{J.~Geophys.~Res.}}    
\def\jqsrt{\ref@jnl{J.~Quant.~Spec.~Radiat.~Transf.}}
\def\memsai{\ref@jnl{Mem.~Soc.~Astron.~Italiana}}
\def\nphysa{\ref@jnl{Nucl.~Phys.~A}}   
\def\physrep{\ref@jnl{Phys.~Rep.}}   
\def\physscr{\ref@jnl{Phys.~Scr}}   
\def\planss{\ref@jnl{Planet.~Space~Sci.}}   
\def\procspie{\ref@jnl{Proc.~SPIE}}   

\newcommand{\half}{\frac{1}{2}}

\newcommand{\mathd}{\ensuremath{\mathrm{d}}}

\newcommand{\both}{\ensuremath{\boldsymbol{\theta}}}

\newcommand{\calP}{\ensuremath{\mathcal{P}}}

\newcommand{\calE}{\ensuremath{\mathcal{E}}}
\newcommand{\calH}{\ensuremath{\mathcal{H}}}

\newcommand{\calG}{\ensuremath{\mathcal{G}}}

\newcommand{\calN}{\ensuremath{\mathcal{N}}}
\newcommand{\fatx}{\ensuremath{\boldsymbol{x}}}

\newcommand{\uk}[1]{\ensuremath{\vec{k}_{\textsc{#1}}}} 
\newcommand{\bk}[1]{\ensuremath{\vec{k}_{#1}}} 

\newcommand{\ri}{\ensuremath{{\rm i}}}

\newcommand{\fus}[1]{\ensuremath{f^*_{#1}}}
\newcommand{\f}[1]{\ensuremath{f(\vec{k}_{#1}})}
\newcommand{\fstar}[1]{\ensuremath{f^*(\vec{k}_{#1}})}

\newcommand{\F}[1]{\ensuremath{f(\vec{k}_{\textsc{#1}}})}
\newcommand{\Fstar}[1]{\ensuremath{f^*(\vec{k}_{\textsc{#1}}})}

\newcommand{\fsub}[1]{\ensuremath{f_{\textsc{#1}}}}
\newcommand{\fsubs}[1]{\ensuremath{f^*_{\textsc{#1}}}}
\newcommand{\psub}[1]{\ensuremath{P_{\textsc{#1}}}}

\title[Edgeworth expansion in cosmology]{On the use of the Edgeworth expansion in cosmology I: how to foresee and evade its pitfalls}

\author[E. Sellentin, A. H. Jaffe, A. F. Heavens]{Elena Sellentin$^{1,2}$, Andrew H. Jaffe$^2$, Alan F. Heavens$^2$
\\
$^{1}$Département de Physique Théorique, Université de Genève, 24 Quai Ernest-Ansermet, CH-1211 Genève, Switzerland\\
$^{2}$Astrophysics Group \& Imperial Centre for Inference and Cosmology, Imperial College London, London SW7 2AZ UK}
\makeatother

\usepackage{babel}
\begin{document}

\date{Accepted 64 AD. Received 1666 AD; in original form 1755 AD}

\maketitle
\pagerange{\pageref{firstpage}--\pageref{lastpage}} \pubyear{2017}

\label{firstpage}
\begin{abstract}
Non-linear gravitational collapse introduces non-Gaussian statistics into the matter fields of the late Universe. As the large-scale structure is the target of current and future observational campaigns, one would ideally like to have the full probability density function of these non-Gaussian fields. The only viable way we see to achieve this analytically, at least approximately and in the near future, is via the Edgeworth expansion. We hence rederive this expansion for Fourier modes of non-Gaussian fields and then continue by putting it into a wider statistical context than previously done. We show that in its original form, the Edgeworth expansion only works if the non-Gaussian signal is averaged away. This is counterproductive, since we target the parameter-dependent non-Gaussianities as a signal of interest. We hence alter the analysis at the decisive step and now provide a roadmap towards a controlled and unadulterated analysis of non-Gaussianities in structure formation (with the Edgeworth expansion). Our central result is that, although the Edgeworth expansion has pathological properties, these can be predicted and avoided in a careful manner. We also show that, despite the non-Gaussianity coupling all modes, the Edgeworth series may be applied to any desired subset of modes, since this is equivalent (to the level of the approximation) to marginalising over the excluded modes. In this first paper of a series, we restrict ourselves to the sampling properties of the Edgeworth expansion, i.e.~how faithfully it reproduces the distribution of non-Gaussian data. A follow-up paper will detail its Bayesian use, when parameters are to be inferred.
\end{abstract}



\section{Scientific Introduction}
The matter fields in the early Universe are compatible with having statistically homogeneous and isotropic Gaussian distributions, according to a series of Cosmic Microwave Background (CMB) experiments \citep{PlanckNG, WMAP}. However, non-linear gravitational collapse then introduces non-Gaussianities into the matter fields, and the effect of these non-Gaussianities on the morphology of the cosmic large-scale structure were phenomenologically described, e.g., by \citet{Coles1, Coles2}. The filamentary structure of the cosmic web, interspersed by voids, is a result of this non-Gaussianity.

If the random fields are described by a set of Fourier modes, then a Gaussian field is characterized by  modes whose phases are independently drawn from uniform distributions, and whose squared amplitudes give the power spectrum. In contrast, non-Gaussianities introduce phase couplings and a coupling between the phases and the absolute values of the Fourier modes. Additionally, the coupling between the absolute values of different Fourier modes leads to the well-known deformation of the non-linear power spectrum on small scales \citep{HaloCamb2,HaloCamb3}. A further effect of the non-Gaussian couplings is that higher-order polyspectra are built up, of which the first two are the bi- and the tri-spectrum. These are the connected three- and four-point functions in Fourier space. A vast literature on cosmological bispectra in the CMB \citep{KSW,primNG,primNG2} or the late large-scale structure (LSS) exists \citep{Fry,MVH,LiciaAndFriends,Bernardeau,Chiaki1,TakJain}.

How a formerly Gaussian random field transforms into a late-time non-Gaussian field due to gravitational processing is analytically not well understood beyond perturbative solutions. Cosmological structure formation is currently best reproduced by sophisticated numerical simulations \citep{Millenium,Illustris,PowSpecStud}, which solve the hydrodynamical equations for a fluid under the influence of gravity. These simulations succeed in producing the pronounced web-like structures as seen on the sky, and can today be used to capture how varying physical effects influence the cosmic web, for example the mass of neutrinos or the finite mean free path of warm dark matter, which smooths out structures on small scales. From a point of view of data analysis, these simulations have however one drawback by construction: they calculate a deterministic evolution of initial conditions. This is because given an initial random field, each simulation solves deterministic differential equations in order to map the initial field to an evolved non-Gaussian field. Due to the deterministic character of the differential equations, N-body simulations therefore do not readily yield a detailed understanding of the random \emph{statistical} changes induced by gravity. Therefore, a direct comparison between data and the theory of structure formation is hindered by a consistent statistical framework, which is able to deal with the difficulties of non-Gaussianity.

The inherent difficulty in non-Gaussian statistics arises because the loss of Gaussianity usually implies a lack of general analytical solutions. Noteworthy exceptions of general applicability are the Edgeworth expansion, which uses cumulants to reconstruct an approximate non-Gaussian distribution function, and DALI \citep{DALI,DALII}, a very general non-Gaussian likelihood approximation which works with quadratic forms rather than cumulants. A numerically-driven method of great generality to describe non-Gaussianities in the cosmological large-scale structure is given in \citet{VIDE}, which detects voids by essentially flooding the cosmic web and thereby detecting voids and their boundaries. Other studies utilize Gaussianization transforms of the cosmological fields, e.g.~\citet{Gaussify1,Gaussify2}.

\begin{figure*}
\includegraphics[width=0.6\textwidth]{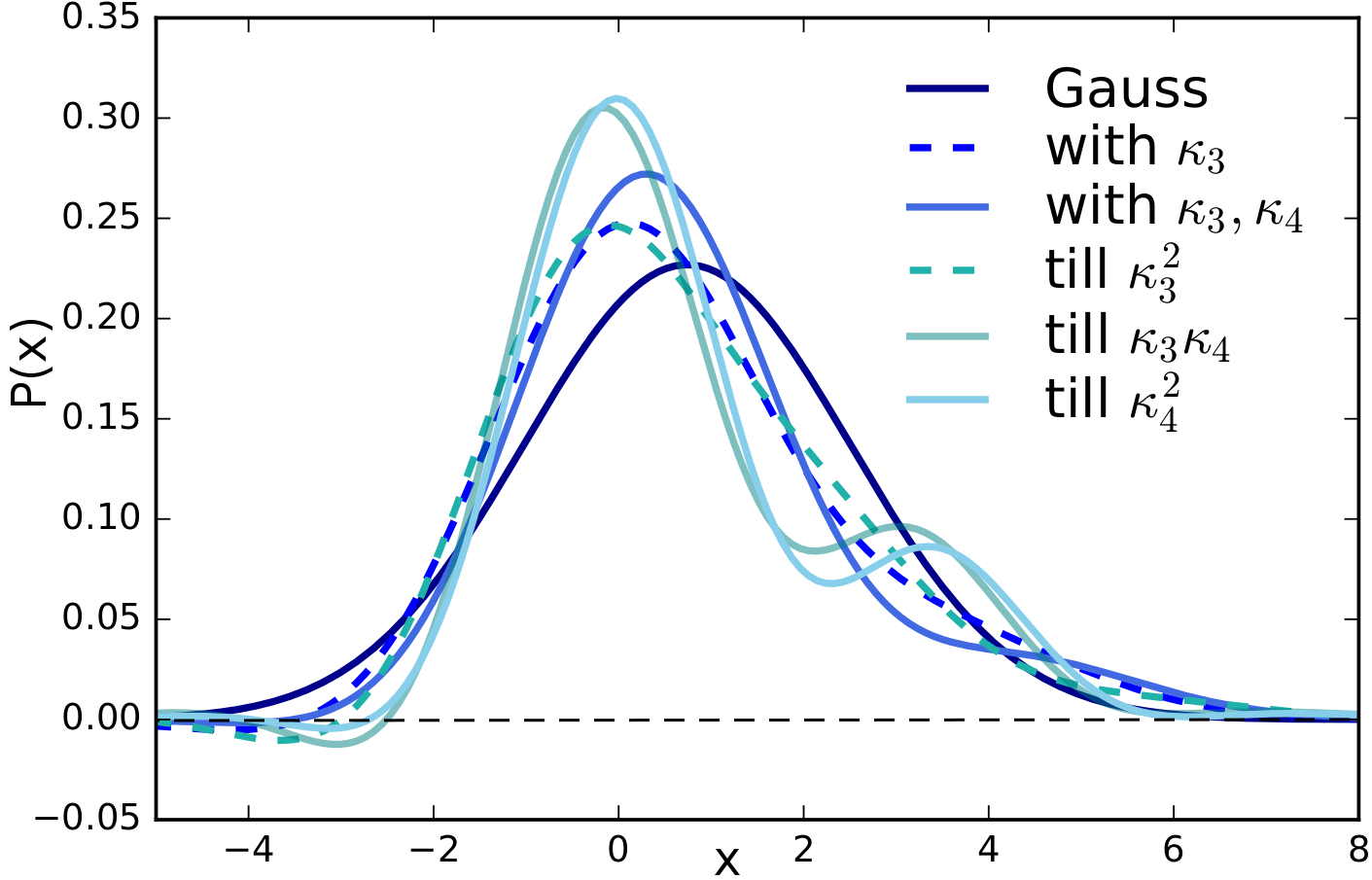}
  \caption{Illustration of the inherent ambiguity when using the Edgeworth expansion to approximate an unknown probability density function: Top to bottom in the legend labels the logical order of including ever higher-order terms in the Edgeworth expansion. Convergence is not achieved by including higher-order terms, because the Edgeworth series is divergent. The figure illustrates the problem of having to judge which of the displayed approximations is the closest to the true probability density function (which is not shown). In this example the true probability density is of course only purposefully blinded, but it would indeed be unknown in a realistic scientific application of the Edgeworth series. We unblind it in Appendix~\ref{Unb}, Fig.~\ref{Fig:Unblind}}
\label{Fig:Ambiguity}
\end{figure*}

Given that current and next generation cosmological surveys \citep{Euclid,Synergies, DES, KiDS} focus on the low-redshift Universe, whose matter fields already exhibit pronounced non-Gaussianities, the interest in describing such non-Gaussianities is increasing in the cosmological community, and hence the interest in the Edgeworth expansion increases alike. This paper therefore focuses on cosmological uses of the Edgeworth expansion. The work at hand extends the framework first described in \citet{Matsu2003} and \citet{Matsu2007}, which also specializes the Edgeworth expansion to Fourier-decomposed fields.

This paper differs from previous analyses, as we intend to clarify the statistical usability of the Edgeworth series in a realistic cosmological setting. We focus especially on how to overcome the problem that the Edgeworth expansion usually introduces unphysical dependencies of the distribution upon its cumulants, which give rise to pathological likelihoods. This is because the expansion is an intrinsically divergent series, in contrast to, e.g., a Taylor approximation. In particular, although the Edgeworth expansion is well known to be an \emph{asymptotic series}, the repercussions of this is not well understood throughout the cosmological community, and here we wish to provide a clear analysis of such mathematical intricacies and how to overcome them. Even though this requires some advanced mathematics,  the effort is worthwhile, since the Edgeworth expansion might be one of very few analytical tools available for the upcoming years.

The problem is illustrated in Fig.~\ref{Fig:Ambiguity}. Provided one true probability density function (which is not shown), six Edgeworth approximations of it were computed. The six approximations differ by including successively higher orders, where the order is given top to bottom in the legend. As can be seen, all six approximations differ in shape and including successively higher-order terms does not improve the convergence towards the true function (not shown). In fact, convergence is not expected from an Edgeworth expansion. However, if the series does not converge, how does one then know which of the given approximations is the best, without knowing the correct distribution? If this question cannot be answered, then the Edgeworth expansion cannot be used to procure a unique and reliable, and especially physically meaningful, probability density function. As the shape of the probability density function of the late cosmic fields is a result of gravitational physics, an ambiguity in the density function would also imply ambiguities in our understanding of gravitational physics, and we hence have to solve this problem. In this paper, we therefore set out to answer the following questions:
\begin{enumerate}
 \item How do we know which of the multitude of possible Edgeworth approximations to a distribution function is best, if the distribution itself is unknown? (And how can `best' be cast into mathematical terms?)
 \item How close is this best approximation to the unknown true distribution?
 \item How strong a non-Gaussianity can we tolerate before the error of the Edgeworth expansion becomes uncontrollable or intolerable?
\end{enumerate}

We shall conclude that the best-known difficulty of the Edgeworth expansion, namely that it can predict negative likelihoods, is the smallest of its problems, and that the problems listed above can be overcome in a systematic and controlled manner.

The outline of this paper is as follows. In order to clarify the necessary mathematical intricacies inherent in non-Gaussian estimation, we restrict the scope of this paper to sampling properties of the Edgeworth expansion. In other words, we shall clarify how reliably it reproduces the distribution of non-Gaussian \emph{data}. Treating this case separately and in exhaustive detail is necessary, because having only one Universe to observe implies that the correct distribution of cosmological large-scale structure data is fundamentally unobservable, which easily leads to misconceptions of the Edgeworth expansion's scope of applicability. Parametric inference with the Edgeworth expansion, where it is regarded as a function of the \emph{parameters} rather than of the data, will be described in the second paper of this series. 

Here, Sect.~\ref{Sect:Ensemble} illustrates the relation between the Edgeworth expansion and the central limit theorem, which is needed to understand the role of Gaussianization by averaging in previous works. Sect.~\ref{1_VolumeFactors} specializes the Edgeworth expansion to Fourier modes of a non-Gaussian field and illustrates the inherent problems in cosmology: we especially discuss what precisely it means if the Edgeworth series is said to `converge' and that this implies cosmological datavectors have to be carefully defined if non-Gaussianities are to not to be destroyed to enforce `convergence'. The upshot of the paper until this point will be that all inherent problems of the Edgeworth expansion are usually addressed by averaging ever more independent random variables, but as this destroys non-Gaussian signals this is exactly what has to be avoided in a cosmological context. The paper hence continues to address the question of how strongly non-Gaussian signals can be treated. Sect.~\ref{Sec:DefD} answers an interesting question of to what extent we can choose the Edgeworth series to describe a  subset of modes, even though the presence of bi- and trispectra now couples all modes. Sect.~\ref{Sec:Div} then solves the issue of ambiguity illustrated in Fig.~\ref{Fig:Ambiguity} by providing a framework which predicts the optimal truncation point for the Edgeworth series, such that the resulting probability density function is closest to the unknown true probability density function. The appendices contain mathematical background material.


\section{Edgeworth expansion and Gaussianization by averging}
\label{Sect:Ensemble}
This section is a short introduction to the Edgeworth expansion and illustrates why its usual setup assumes that one is willing to average non-Gaussianities out. Of course, this is precisely what we will change in the remainder of the paper when we prepare for applying the series to non-Gaussian cosmic fields. Here, we assume the reader has already had some exposure to multivariate statistics, but we provide a pedagogical introduction to multivariate cumulants and their distinction from moments in Appendix~\ref{App:cumulants}, and a pedagogical introduction to the Edgeworth expansion is given in Appendix~\ref{App:EW}.

The Edgeworth expansion is a series approximation to a probability density function, most often expanded around an initial Gaussian distribution. The assumption of expanding around a Gaussian can be dropped without invalidating any of the following derivations -- the only effect would be that more terms appear in the expansion. We use the statistical index notation, where cumulants are indicated by comma-separated indices, and matrix inverses are indicated by pairs of lower indices. Repeated indices on different levels are to be summed over. Natural numbers in square brackets denote the number of permutations. A bold font indicates a vector random variate, and its elements are denoted by a single upper index. For example, $\fatx$ is a multivariate random vector such as a dataset, and $x^i$ is its $i$th element. In contrast, non-random vectors are denoted with arrows, e.g. $\vec{x}$ will in the following be a spatial vector, and $\bk{}$ will be the corresponding wave vector. Using this notation, we have that $\kappa_{i,j}$ is the inverse covariance matrix, and $\kappa^{i,j,\ldots ,k}$ are the $k$th multivariate cumulants. The Edgeworth expansion around a Gaussian is then given by
\begin{equation}
 \calP(\fatx) \approx  \calG(x^i,x^j, \kappa_{i,j})\left[1  + \frac{\kappa^{i,j,k}}{3!}\calH_{ijk}
+ \frac{\kappa^{i,j,k,l}}{4!}\calH_{ijkl}
+ \frac{\kappa^{i,j,k}\kappa^{l,m,n}[10]}{6!}\calH_{ijklmn} + \cdots\right].  
\label{Efirst}
\end{equation}
The tensors $\calH_{ij\ldots k}$ are the multivariate Hermite tensors, where we use the convention Eq.~(\ref{Herms}).

We now briefly recapitulate the connection between the central limit theorem (CLT) and the Edgeworth expansion. To illustrate the argument, it is sufficient to work with the special case of Eq.~(\ref{Efirst}), i.e. a univariate distribution. The CLT states that under mild prerequisites, the average of increasingly many independent random variables follows a Gaussian distribution ever more closely. In the limit of averaging $n \to \infty$ random variables, the Gaussian distribution is recovered. One derivation of the Edgeworth expansion captures the transition region for very large but still finite $n$. This is why the Edgeworth series is often mentioned in the context of `mild' non-Gaussianity, but this mildness is not a necessary prerequisite.

Let $x_i$ be univariate random variables, independent but identically distributed according to the probability density function $\calP(x)$. The distribution $\calP(x)$ is required to have continuous derivatives with respect to the random variable and to be of finite variance, but is otherwise left unspecified. The random samples $x_i$ can be supposed to be mean-subtracted without loss of generality. Otherwise, if the mean is to be kept explicitly, more terms appear in the expansion. All other cumulants $\kappa_2, \kappa_3, \ldots$ are assumed to be finite. We now introduce the \emph{scaled}\footnote{The scaling by $\sqrt{n}$ keeps the width of the recovered Gaussian constant for increasing $n$.} average
\begin{equation}
 Y = \frac{1}{\sqrt{n}} \sum_{i=1}^n x_i,
 \label{Y}
\end{equation}
and study its statistical properties. Its characteristic function is given by the Fourier transform
\begin{equation}
 C_Y(\xi) = \langle e^{ \ri \xi Y}\rangle = \left\langle \exp\left(\ri \xi \frac{1}{\sqrt{n}} \sum_{i=1}^n x_i\right)\right\rangle.
\end{equation}
To exploit the linearity of the average, we now expand the exponential function as a Taylor series, which leads to
\begin{align}
 C_Y(\xi) & =  \sum_{N = 0}^\infty \frac{1}{N!} \left(\frac{\ri\xi}{\sqrt{n}}\right)^N \left\langle \left( \sum_{i=1}^n x_i\right)^N\right\rangle\nonumber\\
 & = 1 + \frac{\ri\xi}{n^{1/2}} \left\langle \sum_{i = 1}^n x_i \right\rangle - \half \frac{\xi^2}{n^{2/2}} \left\langle \left(\sum_{i = 1}^n x_i\right)^2 \right\rangle - \frac{1}{3!} \frac{\ri\xi^3}{n^{3/2}} \left\langle \left( \sum_{i = 1}^n x_i \right)^3 \right\rangle + \cdots + \frac{1}{N!} \frac{(\ri\xi)^N}{n^{N/2}} \left\langle \left( \sum_{i = 1}^n x_i \right)^N \right\rangle + \cdots
\end{align}
Here, we distinguish $n$, the number of samples, from $N$, the index of the $N$th term in the expansion. In the last line, the first averaged term is zero, since it is a multiple of the mean, which we assumed to be zero without loss of generality. For the second average, the square of the sum first has to be expanded, which leads to two types of summands, $x_i^2$ and $2x_ix_j$ with $i \neq j$. Since the samples are however independently drawn, the average $\langle 2 x_i x_j\rangle$ is zero, such that only $n$ terms of $\langle x_i^2 \rangle$ survive. The second term is hence $n\kappa_2(x)$. For the third term, again due to the independence of the samples, averages $\langle x_i x_j x_k\rangle$ will only exist if all indices are equal. The third term is hence $n \kappa_3(x)$, etc. The characteristic function can then be expanded as a power series whose coefficients are proportional to the cumulants,
\begin{equation}
  C_Y(\xi)  = 1 - \frac{\xi^2}{2} \kappa_2(x) - \frac{\ri\xi^3}{3!} \frac{\kappa_3(x)}{n^{1/2}} + \cdots + \frac{(\ri\xi)^N}{N!} \frac{\kappa_N(x)}{n^{N/2-1}} + \cdots,
  \label{sampleEW}
\end{equation}
and the $N$th cumulant will be supressed by powers $n^{N/2-1}$. In the limit $n \to \infty$, all terms apart from the first two will vanish. The first two terms are however the characteristic function of a Gaussian, and we therefore see that the distribution of $Y$ will tend towards a Gaussian  distribution for $n \to \infty$.

The Edgeworth expansion can now be seen from Eq.~(\ref{sampleEW}), when the number of samples $n$ is large but finite. The omitted higher order terms will then be non-zero, but scale as $n^{N/2-1}$. Provided the cumulants $\kappa_N$ do not increase too rapidly with $N$, the non-Gaussian correction terms in the characteristic function will then decrease more or less rapidly. It could hence be hoped that the characteristic function's series expansion can be truncated at a meaningful order $N$ beyond which all corrections can be safely neglected. The Edgeworth expansion then essentially tries to gain an approximate probability density function by backward Fourier-transforming the characteristic function truncated at an order $N$:
\begin{equation}
 \calP(x) \approx \frac{1}{2\pi}\int e^{-\ri\xi x} \left[ 1 - \frac{\xi^2}{2} \kappa_2(x) - \frac{\ri\xi^3}{3!} \frac{\kappa_3(x)}{n^{1/2}} + \cdots + \frac{(\ri\xi)^N}{N!} \frac{\kappa_N(x)}{n^{N/2-1}}\right] \mathd \xi.
 \label{Ctrunc}
\end{equation}
The success of this attempt is however limited by multiple mathematical intricacies like smoothness of derivatives and continuity at the origin. Of these intricacies, two can be singled out as the main reasons why applications of the Edgeworth expansion frequently disappoint: first, even though each individual cumulant can be suppressed by powers of $n^{N/2-1}$, there  exist inifinitely many such higher-order cumulants. In fact, the only distribution known which has a finite number of cumulants is the Gaussian. Including only the first few non-Gaussian cumulants of infinitely many then limits the gained information.

Secondly, the integral over the characteristic function's series will essentially always lead to a divergent series expansion for the probability density function, with the only known counter example given by \citet{CramerConverge}. This means successive higher-order corrections for the probability density increase too fast, such that the approximation does not converge to the target non-Gaussian distribution. The error of $\mathcal{O}(n^{-N/2+1})$ refers only to the precision of the characteristic function when it is truncated after $N$ terms. It is then unrepresentative of the error on the probability density function, whose error can be much larger.

These notions will be made mathematically precise in the following sections, but the derivation above already demonstrates one crucial misconception about what is meant by the `convergence' of the Edgeworth expansion: `convergence' is achieved by increasing $n$, the number of samples, \emph{not} by increasing the number of \emph{terms}. Indeed, when adding more terms, the expansion can develop pathologies, like negative probabilities. In contrast, when truncating the series early and increasing $n$, it behaves ever more soundly and in the end converges towards a Gaussian. In short, the Edgeworth expansion shows the following mathematical behaviour:
\begin{align}
 \textrm{number of samples}~n \to \infty  &\Rightarrow \textrm{convergence (towards a Gaussian)},\\
 \textrm{number of terms}~N \to \infty  &\Rightarrow \textrm{divergence}.
\end{align}
This behaviour is typical for a so-called `asymptotic series', of which the Edgeworth expansion is indeed one example. In summary, the convergence  of an asymptotic series can be enforced when manipulating the series argument, \emph{not} by increasing the number of terms in the sum. This is a crucial difference in comparison to what physicists are used to from, e.g., a Taylor expansion. There, adding higher-order terms improves the convergence towards all the differently shaped functions within the class of analytical functions. For the Edgeworth expansion, the one proven outcome of convergence is always the Gaussian distribution. This implies that usual applications of the Edgeworth expansion for a non-Gaussian probability distribution work by averaging the non-Gaussianity out. This is also the case for the applications presented in \citet{Matsu2007}.

In a cosmological application where the non-Gaussianity is the signal of interest, averaging the non-Gaussianity away is obviously counterproductive. In the following sections, we will hence first analyse why previous applications of the Edgeworth expansion to Fourier-decomposed fields reduce the inherent non-Gaussianities, and we will then come back at the asymptotic properties of the Edgeworth expansion and exploit them in order to investigate how averaging non-Gaussianities out can be overcome, such that strong non-Gaussianities can be investigated instead.


\section{The Edgeworth expansion in the Fourier domain --- the role of volume factors}
\label{1_VolumeFactors}
In \citet{Matsu2007},  the Edgeworth expansion was used to obtain an approximate probability density function for the Fourier modes of a random field\footnote{Note that \citet{Matsu2007} uses the cumulant expansion theorem to derive the series, but the series is not referred to as Edgeworth.}. We give here a somewhat shorter rederivation of the mathematics. The aim is to clarify the meaning of volume factors which appear in \citet{Matsu2007} as these have triggered further interest in the community \citep{Aegg,WCO}. We shall especially see why and how these volume factors relate to convergence by reducing non-Gaussianities.

In comparison to \citet{Matsu2007}, we use the opposite sign-convention in the Fourier phase, and we will furthermore underline the independence of the results from the convention of the Fourier transform. Let $f(\vec{x})$ be a random field in real-space, then we define its Fourier transform to be
\begin{align}
 f(\vec{k}) = \int f(\vec{x}) e^{\ri \bk{} \cdot \vec{x}} \mathd^d x,
 \label{for}
\end{align}
where $d$ is the spatial dimension, so typically $d=2$ for a survey at fixed redshift or in a narrow shell, or for a small patch of the cosmic microwave background. For $d=3$, a cosmological volume is considered instead.

The corresponding backwards transform is given by
\begin{align}
 f(\vec{x}) = \frac{1}{(2\pi)^d}\int f(\vec{k}) e^{-\ri \bk{} \cdot \vec{x}} \mathd^d k.
 \label{back}
\end{align}
Independent of any convention, the reality of the real-space fields leads to the Hermitian redundancy
\begin{equation}
 \f{i} = f^*(-\vec{k}_i),
 \label{HR}
\end{equation}
if we denote complex conjugation with an asterisk.
Depending on our chosen convention, the second cumulant in Fourier space is the power spectrum
\begin{equation}
\kappa^{i,j} = \langle \f{i}\f{j}\rangle = (2\pi)^d \delta(\bk{i} + \bk{j}) P(k_i)\;,
\label{kappa2}
\end{equation}
where the Dirac delta function results from the assumption of statistical homogeneity, i.e., that the configuration space correlation function satisfies $\left<f(\vec{x}_1)f(\vec{x}_2)\right>=\xi(|\vec{x}_1-\vec{x}_2|)$, and that the power spectrum $P(k)$ is the Fourier transform of $\xi(\vec{x})$, and we have assumed isotropy in writing $P$ as a function of $k\equiv |\vec{k}|$. Note that the presence of the delta function in this definition applies for the continuum limit of the Fourier transform, and as usual is only well-defined inside an integrand.  We shall in practice be dealing with discrete modes, for which the orthogonality is expressed with a Kronecker delta. Summations in the discrete case with a Kronecker delta become integrals with a Dirac delta function in the continuum limit.

If we do indeed define our Fourier transform over a finite volume $V=L^d$,
\begin{align}
 f_V(\vec{k}) = \frac1V \int_V f(\vec{x}) e^{\ri \bk{} \cdot \vec{x}} \mathd^d x,
 \label{forV}
\end{align}
with inverse given by the sum
\begin{equation}
    f(\vec{x}) = \sum_{\vec{k}} f_V(\vec{k})e^{-\ri \bk{} \cdot \vec{x}},
\end{equation}
taken over the usual discrete ${\vec{k}}$ with spacing $\delta k=2\pi/L$ in each direction. Note that this convention is such that $f(\vec{x})$ and $f_V(\vec{k})$ have the same units. In this case, the second order cumulant is
\begin{equation}
\langle f_V(\vec{k}_i) f_V(\vec{k}_j)\rangle = P(\vec{k}_i)\delta_{i,j}\;,
\end{equation}
where $\delta_{i,j}$ is the Kronecker delta, and the power spectrum $P(\vec{k})$ is again given by the Fourier transform (with these conventions and over the domain $V$) of the real-space correlation function, which is independent of Fourier convention, and determined by the underlying theory. Note that $P=V^{-1}\int_V \xi(\vec{r})e^{\ri \bk{}\cdot \vec{r}} \mathd^d x$ differs by a factor $V$ from the $P$ in the continuous case, Eq.~(\ref{kappa2}).

This generalizes to higher-order correlation functions only being functions of the shape of the polygon spanned by the $\vec{x}_i$, rather than their overall orientation.
Since complex conjugation flips the sign of the $\vec{k}$-vectors, the power spectrum can equivalently be written as
\begin{align}
\langle \f{i}\fstar{j}\rangle & = (2\pi)^d \delta(\bk{i} - \bk{j}) P(k_i).
 \label{kappa2star}
\end{align}
The third cumulant is given by the bispectrum $B(k_i,k_j,k_k)$, defined by
\begin{equation}
 \kappa^{i,j,k} = \langle \f{i}\f{j}\f{k}\rangle = (2\pi)^d \delta(\bk{i} + \bk{j} + \bk{k}) B(k_i,k_j,k_k),
 \label{kappa_3}
\end{equation}
where all $\bk{i}$ appear with a positive sign in the delta function since none of the Fourier modes have been conjugated. An example for such a bispectrum is the tree-level bispectrum due to gravitational collapse as given by \citep{Fry,Bernardeau}
\begin{align}
B(\bk{1},\bk{2},\bk{3}) =  2 F_2(\bk{1},\bk{2})P(k_1)P(k_2) [3],
\end{align}
recalling that $[3]$ refers to a sum over expressions of this form with the three permutations of the indices $\{1,2,3\}$, and with
\begin{equation}
 F_2(\bk{1},\bk{2}) = \frac{5}{7} + \frac{\bk{1}\cdot \bk{2}}{2 k_1 k_2} \left( \frac{k_1}{k_2} + \frac{k_2}{k_1}\right) + \frac{2}{7}\left(\frac{\bk{1}\cdot \bk{2}}{k_1 k_2} \right)^2.
\end{equation}
Up to third order, moments and cumulants are identical; only from fourth order onwards do they  differ by disconnected terms. The fourth cumulant is given by the connected four-point function in Fourier space and is known as the trispectrum, $T$, defined by
\begin{equation}
 \kappa^{i,j,k,l} = \langle \f{i}\f{j}\f{k}\f{l}\rangle_\textrm{c} = (2\pi)^d \delta(\bk{i} + \bk{j} + \bk{k} + \bk{l}) T(\bk{i},\bk{j},\bk{k}, \bk{l}),
 \label{kappa4}
\end{equation}
where the subscript `c' refers to the connected (non-Gaussian) part of the correlation.
Calculating a theoretical prediction for the trispectrum of evolved density fields is currently at the edge of our capabilities as a scientific community. Halo-model predictions and response approaches have however recently been explored, especially for the calculation of covariance matrices of evolved matter fields, for example by \cite{KiDS,TakJain,Barreira}. In total, given that perturbative predicitions for polypsectra of evolved density fields exist \citep{PlanckNG,Fry,Bernardeau}, and given that improvement can be expected in the upcoming years, one could hope that knowing these cumulants immediately leads to clear information on the non-Gaussian probability density function. This is however not the case because simply plugging the bispectrum into the Edgeworth expansion leads to ambiguous probability density functions, which may contain more pathologies than actual physical information (as illustrated in Fig.~\ref{Fig:Ambiguity}). The upcoming derivations will hence detail on how much physical information about the probability density function can be gained with an Edgeworth expansion, and how this depends on the definition of the data vector, and the truncation order of the series.

Apart from the question what physical information is contained in the Edgeworth expansion, this physical information must additionally be independent of the convention chosen for the Fourier transform.
We have so far defined the transforms over both an infinite volume and a finite domain $V=L^d$, with the latter giving an infinite but discrete spectrum of modes with spacing $\Delta k=2\pi/L$. Numerical implementations of a Fourier transform are additionally restricted to a finite number $s$ of discrete samples from a signal and hence do not only contain volume factors, but also the number of these samples. The location of the actual factors $V$ and $s$ in the definition of the transform and its inverse depends on the Fourier conventions. However, these factors simply serve the purpose to make the Fourier basis not only orthogonal but also ortho\emph{normal} (where the normalization is to a Dirac delta function
in the case of infinite volume).

We hence generalize the notation for the polyspectra as follows. We imagine $f(\vec{k}_i)$ are Fourier modes obtained in any convention, and any domain. Both the convention and the domain will determine the numerical value and the dimension of these Fourier modes. Such changes in the Fourier modes will then automatically propagate into expectation values of their $N$-point functions.

We define for the bispectrum
\begin{align}
 \kappa^{i,j,k} & = \langle f(\vec{k}_i) f(\vec{k}_j)f(\vec{k}_k)\rangle  \nonumber \\
 & = {\tilde B}(  \vec{k}_i, \vec{k}_j,\vec{k}_k  ) \delta^i_{j+k} \nonumber \\
 & = B^{i,j,k},
\end{align}
and for the trispectrum
\begin{align}
 \kappa^{i,j,k,l} & = \langle f(\vec{k}_i) f(\vec{k}_j)f(\vec{k}_k)f(\vec{k}_l)\rangle  \nonumber \\
 & = {\tilde T}(\vec{k}_i, \vec{k}_j, \vec{k}_k,\vec{k}_l) \delta^i_{j+k+l} \nonumber \\
 & = T^{i,j,k,l},
\end{align}
where the middle line in each of these equations defines bi- and tri-spectra, ${\tilde B}$ and ${\tilde T}$ which still have to be multiplied by a delta function to enforce closed polygons,  $\sum_i \vec{k}_i=0$, and the final line is a tensor which includes both the polyspectrum shape and the delta factor.

In this notation, the likelihood for Fourier modes of a Gaussian random field is as follows: the real parts $a(\bk{i})$ and imaginary parts $b(\bk{i})$ of a single Fourier mode are independently drawn from univariate Gaussian distributions.  In the case of discrete $\vec{k}$ modes, we have
\begin{equation}
 a(\bk{i}) \sim \frac{1}{\calN} \exp\left(-  \frac{a^2(\bk{i})}{P^i} \right),
\end{equation}
and
\begin{equation}
 b(\bk{i}) \sim \frac{1}{\calN} \exp\left(-  \frac{b^2(\bk{i})}{P^i} \right),
\end{equation}
where $P^i\equiv P(k_i)$ and $f(\vec{k}_i)f(\vec{k}_i)$ have the same dimension and the argument of the exponential is hence dimensionless, as required.

The likelihood using the Fourier modes themselves is then given by
\begin{align}
\calG \left( f(\vec{k}_i)  \right) \propto
\exp \left( -  \frac{f(\vec{k}_i)f^*(\vec{k}_i)}{P(k_i)} \right).
\end{align}

Equipped with these results, we can now derive the Edgeworth expansion for the Fourier modes. The derivation of the Edgeworth series in Sect.~\ref{Sect:Ensemble} was abandoned at Eq.~(\ref{Ctrunc}), where we had only mentioned that the expansion attempts to solve the backwards Fourier transform of the truncated characteristic function. The next steps in the derivation are given in Appendix~\ref{App:EW}, and one finally arrives at the multivariate Edgeworth expansion
\begin{equation}
\calP (\fatx) \approx \left[ 1 - \frac{\kappa^{i,j,k}}{3!} \partial_i\partial_j\partial_k + \frac{\kappa^{i,j,k,l}}{4!} \partial_i \partial_j \partial_k \partial_l + \frac{\kappa^{i,j,k}\kappa^{l,m,n}}{6!} \partial_i\partial_j\partial_k\partial_l\partial_m\partial_n [10] + \cdots\right]\mathcal{G}(x^i,x^j, \kappa_{i,j}).
 \label{OperatorVersion}
\end{equation}
The partial derivatives now give rise to Hermite tensors, which are the multivariate extensions of the Hermite polynomials. The multivariate Edgeworth expansion then reads
\begin{equation}
\calP(\fatx) \approx  \calG(x^i,x^j, \kappa_{i,j})\left[1 + \frac{\kappa^{i,j,k}}{3!}\calH_{ijk}
+ \frac{\kappa^{i,j,k,l}}{4!}\calH_{ijkl}
+ \frac{\kappa^{i,j,k}\kappa^{l,m,n}[10]}{6!}\calH_{ijklmn} + \cdots \right]  
 \label{hermiedge}
\end{equation}
with the Hermite tensors as defined in Appendix~\ref{App:Hermi}.

To apply the Edgeworth expansion to Fourier decomposed fields, we now have to extend the expansion from real- to complex-valued data. As Fourier modes of a real-valued field satisfy the Hermitian redundancy Eq.~(\ref{HR}), only half of the Fourier modes are independent and need to be used in the data vector. We therefore split the set of all Fourier modes $\{ f(\bk{i}) \}$ into two subsets, where each subset never contains both a Fourier mode $f(\bk{i})$ and its Hermitian partner $f(-\bk{i})$, but it is otherwise irrelevant how the Fourier modes are distributed onto the two sets. We assume the field under consideration has been mean-subtracted, and hence exclude the Fourier mode $f(\vec{0})$. Clearly, only one of the two subsets is needed to form a data vector, if the Hermitian redundancy is additionally enforced -- if the Hermitian redundancy is not enforced, all Fourier modes have to be used as data vector. We hence introduce the additional notation that $\f{i}$ runs over both subsets, i.e., includes all Fourier modes, whereas $\F{i}$ with capital indices only runs over one of the subsets, thereby excluding all modes which are redundant due to the Hermiticity; this is equivalent to the sums over the `upper half sphere' of modes in \citet{Matsu2007}.

We consider it easiest to derive the Edgeworth expansion for Fourier modes by beginning all calculations in the complex plane. The likelihood using the Fourier modes from the upper half sphere is given by
\begin{align}
\calG \left( \{f(\uk{i}) \} \right) = \frac{1}{\calN'} \prod_I \exp \left( -  \frac{\F{i}\Fstar{i}}{P(k_I)} \right).
\label{Pmodes}
 \end{align}

In the continuous (Fourier transform) case, we may proceed formally by defining the inverse correlation function $\xi^{-1}_k$ in Fourier space as in \cite{Matsu2007}, Eq.~(9) such that
\begin{equation}
\int d^d\vec{k}'' \,\xi_k(\vec{k}, \vec{k}'') \,\xi^{-1}_k(\vec{k}',\vec{k}'') = \delta(\vec{k}-\vec{k}')
\end{equation}
where $\xi_k(\vec{k}, \vec{k}')=\langle f(\vec{k})f(\vec{k}')\rangle$, from which it follows that the inverse correlation matrix is given by $\kappa_{i,j} = (2\pi)^{-d}(P^i)^{-1} \,\delta(\vec{k}_i+\vec{k}_j)$.  Equation~(\ref{Pmodes}) then generalises formally to
\begin{align}
\calG \left( \{f(\uk{i}) \} \right) \propto \exp \left( -  \int \frac{\F{I}\Fstar{I}}{P(k_I)}\frac{d^d{\vec{k}_I}}{(2\pi)^d} \right).
\label{PmodesFT}
\end{align}

The Edgeworth expansion of a non-Gaussian distribution for evolved matter fields now requires the complex derivatives of the initial Gaussian distribution. The real and imaginary parts could be treated as independent variables, but the calculations are greatly facilitated if the complex Fourier modes themselves are used as the data. The derivative of the Gaussian with respect to the complex Fourier modes is
\begin{equation}
 \frac{\partial \calG}{\partial \fsub{i}} = \half \left( \frac{\partial \calG}{\partial a_\textsc{i}} - \ri \frac{\partial \calG}{\partial b_\textsc{i}} \right)
\end{equation}
and
\begin{equation}
 \frac{\partial \calG}{\partial \fsubs{i}} = \half \left( \frac{\partial \calG}{\partial a_\textsc{i}} + \ri \frac{\partial \calG}{\partial b_\textsc{i}} \right)
\end{equation}
because $(\partial \calG / \partial \fsub{i})^* = \partial \calG/\partial\fsubs{i}$ since $\calG$ is real-valued. In contrast, the derivatives of the complex-valued Fourier modes with respect to their conjugate vanishes ${\partial \fsub{i}}/{\partial \fsubs{i}} = 0$.
For example, the first derivative of the Gaussian is
\begin{align}
 &\frac{1}{\calG} \frac{\partial \calG}{\partial \fsub{i}} = -\frac{\fsubs{i}}{\psub{i}}\nonumber\\
 &\frac{1}{\calG} \frac{\partial \calG}{\partial \fsubs{i}} = - \frac{\fsub{i}}{\psub{i}},
\end{align}
and the distinct second derivatives,
\begin{align}
 \frac{1}{\calG} \frac{\partial^2 \calG}{\partial \fsub{i} \fsub{J}} &= \frac{\fsubs{i}\fsubs{j}}{\psub{i}\psub{j}} \nonumber\\
 \frac{1}{\calG} \frac{\partial^2 \calG}{\partial \fsub{i} \fsubs{J}}& = \frac{\fsubs{i}\fsub{j}}{\psub{i}\psub{j}} - \frac{\delta^I_J}{\psub{i}}.
\end{align}
The Edgeworth expansion is then
\begin{equation}
\left[1 - \frac{\kappa^{i,j,k}}{3!}\frac{\partial}{\partial f_i} \frac{\partial}{\partial f_j} \frac{\partial}{\partial f_k}  + \frac{\kappa^{i,j,k,l}}{4!}\frac{\partial}{\partial f_i} \frac{\partial}{\partial f_j} \frac{\partial}{\partial f_k} \frac{\partial}{\partial f_l} + \cdots\right]     \calG \left( \{f(\vec{k}) \} \right).
\end{equation}
Similarly, the higher order terms can be worked out and then be contracted with the cumulants, which are given by the polyspectra Eqs.~(\ref{kappa2}--\ref{kappa4}). These contractions give us a final equation for the multivariate Edgeworth expansion, the first few terms of which are
\begin{align}
\calP[\{ f(\vec{k})\}] = \calG(\{ f(\vec{k}) \}) \cdot\bigg[ 1 + &
\frac{B^{i,j,k}}{3!} \frac{\fus{i} \fus{j}\fus{k}}{P_iP_jP_k} \nonumber\\
 + & \frac{T^{i,j,k,l}}{4!} \left(
\frac{\fus{i}\fus{j}\fus{k}\fus{l}}{P_iP_jP_kP_l} - \frac{\fus{i}\fus{j}
\delta^k_l }{P_i P_jP_k} [6] + \frac{\delta^i_j \delta^k_l}{P_i P_k}[3] \right) \nonumber\\
+ & \frac{B^{i,j,k} B^{l,m,n}[10]}{6!} \left(
\frac{\fus{i}\cdots\fus{n}}{P_i\cdots P_n} \right. -
\frac{\fus{i}\fus{j}\fus{k}\fus{l} \delta^m_n}{P_i
P_j P_k P_lP_n}[15] \nonumber\\
&\phantom{\frac{B^{i,j,k} B^{l,m,n}[10]}{6!} \bigg(}   
+ \frac{\fus{i}\fus{j} \delta^k_l
\delta^m_n}{P_i P_j P_k P_m}[45]\nonumber\\
& \phantom{\frac{B^{i,j,k} B^{l,m,n}[10]}{6!} \bigg(}    
- \left.\frac{\delta^i_j
\delta^k_l \delta^m_n}{P_i P_k
P_l}[15]  \right)\nonumber\\
+ & \cdots \bigg]\;,
\label{fullEW}
\end{align}
where subscripts $i,j,\ldots$ on $f$ or $P$ refer to evaluation of the quantity at $k_i, k_j, \ldots$, and all indices inside the square brackets are to be contracted away. As written here, the Edgeworth expansion can refer to the distribution of the full set of independent modes $\left\{f(\vec{k}_I)\right\}$ or any subset thereof by selecting the appropriate modes on both the right and left sides of Eq.~\ref{fullEW} (see Sect.~\ref{Sec:DefD}).
This expression can be compared to Eq.~(48) of \citet{Matsu2007}. That work uses a convention for Fourier transforms and polyspectra in which the cumulants of a field decomposed via a discrete (finite volume) Fourier transform are
\begin{align}
 \langle f_{\bk{1}}\cdots f_{\bk{N}}\rangle_c & = V^{1-N/2} \delta^K_{\bk{1} +\cdots+\bk{N}} P^{(N)}(\bk{1},\ldots ,\bk{N-1})\nonumber\\
 & = \kappa^{i,j,\ldots ,N},
 \label{Vapp}
\end{align}
where the last line establishes the connection with the notation used in our paper. The volume $V$ is that of the finite region that is Fourier transformed, and, in combination with the Kronecker delta factors, replaces the Dirac delta function in the infinite volume case. In Matsubara's convention, the higher-order cumulants (the polyspectra) hence scale with the volume as
\begin{align}
\kappa^{i,j} & \propto V^{0}\;,\\
 \kappa^{i,j,k} & \propto V^{-1/2}\;,\\
 \kappa^{i,j,k,l} & \propto V^{-1}\;,\ldots
\end{align}
and similarly for higher orders. This is akin to the $N$th higher-order cumulant being suppressed by factors $n^{-N/2+1}$, when the ensemble average of $n$ independent samples is taken. \citet{Matsu2007} discusses that taking the volume \emph{formally} to infinity makes the series approximation `converge', and that furthermore the series should be dominated by terms of lower order in $1/V$. However, the exact mathematical sense in which this convergence is meant is omitted. In fact, following this argument, the Gaussian likelihood is regained for $V\to \infty$, indicating that the joint probability density function of Fourier modes of a non-Gaussian field allegedly tends to a Gaussian if the field is surveyed in large enough a volume. This is in itself a misleading statement, and can easily be interpreted incorrectly. In \citet{Matsu2007}, the author seems to be aware of it, as he later on constructs examples which evade the confusing part of this statement. We here wish to present clearly in which sense the $V\to\infty$ limit can lead to a Gaussian distribution, and why this is insufficient if non-Gaussian cosmological density fields are to be described.
Essentially, all applications in which $V\to \infty$ leads to a Gaussianization imply that physically existing non-Gaussianity is averaged away due to the definition of various estimators. We shall now investigate this issue in more detail.

We saw in Sect.~\ref{Sect:Ensemble} that ensemble averages over random samples lead to a suppression of higher-order cumulants, such that the joint distribution of averaged samples tends to a Gaussian distribution for $n\to \infty$, if $n$ is the number of samples. The role of $n$ is now taken by $V$, which is essentially equivalent to the assumption of ergodicity: subvolumes take the role of independent realizations. The statement that the statistics of non-Gaussian fields tends back towards Gaussianity when the field is observed in a sufficiently large volume, is --- as such --- incorrect. It depends on how the data vector is defined and how it is actually determined from the real-space field.

Cosmologists familiar with bandwidth-type estimators for the power spectrum or bispectrum know that for $V\to\infty$, ever more Fourier modes of similar $|k|$ will be available. For example, let $S_k \approx 4 \pi k^2 \Delta k$ with $\Delta k /k \ll 1$ be the shell of Fourier modes with equal $|k|$ but different orientations $\bk{}$. In the discrete Fourier transform, this shell will be populated with a number of modes proportional to the density-of-states $g = V/(2\pi)^d$. If the survey volume $V$ increases, the shell will therefore be ever more densely populated. If we denote the number of vectorial Fourier modes in the shell by $N_k = g S_d(k)\Delta k$, where $S_d(k)$ is the surface area of a $d-$sphere of radius $k$,  then the usual power spectrum estimator is
\begin{equation}
 \hat{P}(k) = \frac{1}{N_k} \sum_{i = 1}^{N_k} f(\bk{i})f^*(\bk{i}).
\end{equation}
Hence for increasing $V$, the sum runs over more terms, and the power spectrum estimator becomes less noisy. For $N_k \to \infty$, the distribution of this power spectrum estimator tends to a Gaussian, but in general 2-point correlation functions estimated in this manner follow gamma-type distributions \citep{Hami,BJK,SH_mat}. This is not to be confused with the underlying \emph{field} following a Gaussian distribution, because obviously Gaussian \emph{and} non-Gaussian fields both have power spectra, but the non-Gaussian fields have additional polyspectra. In fact, a similar estimator for the bispectrum is
\begin{equation}
 \hat{B}(k_1, k_2, k_3) = \frac{1}{N_t} \sum_{i = 1}^{N_t} [f(\bk{1})f(\bk{2})f(\bk{3})]_i,
\end{equation}
where $N_t$ is the number of closed triangle configurations, where each triangle has side lengths $k_1,k_2$ and $k_3$. Again, for increasing $V$ more such triangles will be available, where the exact number of each triangle type depends on the triangle shape. In total however, the bispectrum estimator becomes again less noisy, thereby proving that for this type of estimator, the non-Gaussianity of the field does not disappear in the limit $V\to\infty$, but rather can be measured ever more accurately. The same holds true for higher order polyspectra, proving that the inherent non-Gaussianity of a field does not disappear if the survey volume is increased.

The non-Gaussianity that can disappear if the volume is increased refers to another estimation technique. Let $\hat{f}(\bk{})$ be the realization of a Fourier mode within a volume $V_s$. The modulus $|k|$ and direction $\vec{k}$ are to be kept fixed. Let now $V \gg V_s$ be a volume into which $V_s$ can be fitted $n$ times. If one is willing to ignore statistical correlations between the subvolumes of a field (e.g. if one observes an infinite volume one could single out infinitely many subvolumes with infinite separations), then one can argue that each sub-volume is an independent realization of the non-Gaussian field. For the sake of simplicity (but this also holds in the general case), we furthermore assume the regarded Fourier mode does not couple to other Fourier modes. Within each subvolume, the Fourier mode $\hat{f}(\bk{i})$ is hence drawn from the same statistical distribution with cumulants $\kappa_2, \kappa_3, \kappa_4, \ldots$. We can then define the quantity
\begin{equation}
 f(\vec{k}) = \frac{1}{\sqrt{n}} \sum_{i = 1}^n [\hat{f}(\bk{})]_i,
\end{equation}
which averages the Fourier mode between the subvolumes. Then, the argument of Sect.~\ref{Sect:Ensemble} holds true, and the higher-order cumulants will decrease with powers of $n^{N/2-1}$. Since the number of sub-volumes of fixed size scales as $n\propto V$ if the volume of the survey is increased, this explains why \citet{Matsu2007} finds that for $V\to\infty$, the non-Gaussianity disappears.

Upon closer inspection, this estimation procedure is however highly dissatisfactory: since the Fourier mode is initially drawn from a distribution with zero mean, averaging for $V\to \infty$ will make the Fourier mode disappear. In fact, the entire structure of the random field will then disappear, and one is left with a constant background. This estimation procedure is therefore of no practical relevance in cosmology even though it is mathematically true that for $V\to\infty$ one regains a Gaussian distribution.

Another possibility would be to try and estimate the average modulus of a Fourier mode by averaging over sub-volumes. This estimator would read
\begin{equation}
 |f(\vec{k})| = \frac{1}{\sqrt{n}} \sum_{i = 1}^n [|\hat{f}(\bk{})|]_i.
\end{equation}
In the limit $V\to\infty$, this estimator does not tend towards zero. This is indeed an example \citet{Matsu2007} describes in his Section VI\footnote{Note, that in his Fig.~3 and Fig.~8, the author writes `Gaussian' where instead `Rayleigh'-distribution is meant.}.

A further example, in which the increase of survey volume can lead to an effective Gaussianization, is when the statistics of localized objects is regarded. This may apply to quasars, or galaxy clusters, as long as it can be assumed that the number density of such objects is spatially constant. Then the number $n$ of such objects will increase linearly with the survey volume $V$, such that again, the increasing volume implies averaging more independent draws from an unobserved non-Gaussian ensemble.

In total however, the Edgeworth expansion as so far presented cannot be used to accurately describe the non-Gaussian distribution function of Fourier modes of the late cosmic fields. To be precise, the problem occurs because for $V\to\infty$ the accuracy of the distribution improves by destroying non-Gaussianity, i.e. the estimated higher order cumulants tend to zero  e.g. for the bispectrum
\begin{equation}
 V\to\infty \ \ \ \Rightarrow \ \ \ \hat{B}(k_1, k_2, k_3) \to 0,
\end{equation}
whereas we of course wish that for increasing volume, we can \emph{better} describe the non-Gaussianity, i.e. we target
\begin{equation}
\ \ \ \ \ \ \ \ \ \ \ \ V\to\infty \ \ \  \Rightarrow \ \ \ \hat{B}(k_1, k_2, k_3) \to B^{\rm theo}(k_1, k_2, k_3).
\end{equation}
But if we are not willing to destroy non-Gaussianities in order to improve how accurately the Edgeworth expansion describes the probability density of the data, then we have to provide further justification for how we define our data vector, and how we truncate the series. These problems will be addressed in the next sections.


\section{Truncation order and choice of modes}
\label{Sec:DefD}

We envisage using the Edgeworth series to approximate the yet unknown probability density function of Fourier modes of a non-Gaussian field. This aim raises the question of how the accuracy of such an approximation can be monitored, and what it depends on. Since the accuracy of the Edgeworth expansion is in usual setups improved upon by averaging the non-Gaussianity away, whereas we wish to precisely avoid destroying the non-Gaussianity, we have to create novel solutions for monitoring the accuracy. The following conceptual problems have to be addressed.

We have to understand how to truncate the divergent Edgeworth series meaningfully, since it provides a multitude of differently shaped approximations $\calP_N(\fatx)$, depending on the truncation order $N$. An example is shown in Fig.~\ref{Fig:Ambiguity}, where it is unclear which of the plotted approximations represents the unknown true probability density function best. This illustrates the well-known problem that even infinitely many moments and cumulants do not specify a probability density function uniquely. From this set of all possible approximations, we hence have to single out the best one, which arises from the optimal truncation order $N$ for the Edgeworth series. We then need to know how accurately this approximation represents the true probability density function. As the true non-Gaussian distribution is unknown, predicting the accuracy needs to be achieved without referring to the unknown function. In comparison to the difficulties due to the divergence of the series and non-uniqueness of its approximations, the fact that the Edgeworth expansion can predict negative probabilities is a minor nuisance.

A further issue is that the Edgeworth expansion provides an approximation for the sampling distribution of a given data vector - i.e. some \emph{finite} subset of infinitely many modes, all of which will be correlated. This raises the question of whether selecting only a finite number of modes is self-consistent.  The point is that higher-order correlations couple modes together, so it is not obvious that one could discard some set of modes, and consider the sampling distribution of the remainder in isolation. We therefore begin by demonstrating that in fact we are allowed to choose any finite-dimensional data vector, and the Edgeworth expansion will approximate its distribution, even if correlations with neglected data points exist.

If a finite-dimensional data vector is to be treated self-consistently within a likelihood approach, then this requires the likelihood to have sound marginals with respect to all omitted data points. Therefore, if $L$ is the likelihood of data points $x^i$, then we require that
\begin{equation}
\int L(x^i,x^j,\ldots ,x^{n-1},x^n, x^{n+1},\ldots ,x^z) \mathd x^n {=} L(x^i,x^j,\ldots ,x^{n-1},x^{n+1},\ldots ,x^z),
\label{margcond}
\end{equation}
meaning the dimensional reduction by marginalizing data point $x^n$ must reproduce the same result as if the data point $x^n$ had been omitted. If Eq.~(\ref{margcond}) is fulfilled by a likelihood, then data points omitted from an analysis are implicitly marginalized over. Not all probability density functions automatically satisfy the condition Eq.~(\ref{margcond}) and can therefore not be used as sound likelihoods. However, the Edgeworth expansion does approximately satisfy Eq.~(\ref{margcond}), as we now show.

By definition, the characteristic function $C_x(\boldsymbol{\xi})$ is the Fourier transform of a probability density $\calP(\fatx)$. The joint distribution of two data points is hence
\begin{equation}
 \calP(x^i, x^j) = \frac{1}{(2\pi)^2} \int C_x(\xi^i, \xi^j) \exp\left(-\ri \xi^i x^i \right) \exp\left(-\ri\xi^jx^j\right) \mathd \xi^i \mathd \xi^j.
\end{equation}
The marginal likelihood of $x^i$ is defined as
\begin{equation}
 \calP(x^i) = \int \calP(x^i, x^j) \mathd x^j.
\end{equation}
We can hence insert the characteristic function and integrate it over $x^j$. The only $x^j$ dependence is however in the second Fourier phase, and therefore
\begin{align}
 \calP(x^i) & = \frac{1}{(2\pi)^2} \int C_x(\xi^i, \xi^j) \exp\left(-\ri \xi^i x^i \right) \mathd \xi^i \mathd \xi^j \int \exp\left(-\ri\xi^j x^j\right)  \mathd x^j \nonumber\\
 & = \frac{1}{(2\pi)^2} \int C_x(\xi^i, \xi^j) \exp\left(-\ri \xi^i x^i \right) \mathd \xi^i \mathd \xi^j (2\pi) \delta(\xi^j)\nonumber \\
 & = \frac{1}{2\pi} \int C_x(\xi^i, 0) \exp(-\ri\xi^i x^i) \mathd \xi^i.
\end{align}
The last line is indeed the Fourier transform of the marginal density $\calP(x^i)$, so provided the integral exists, data points can be marginalized by omitting their Fourier counterparts from the characteristic function. This is indeed akin to simply omitting a data point altogether, and hence proves that we can self-consistently select a finite range of modes out of infinitely many correlated modes.

The above holds true, because the Fourier phase for $x^j$ was integrated over to yield the delta-function. The Edgeworth approximation now however avoids exactly this integral. Additionally, it also does not use the full characteristic function, but its truncated series expression instead. Therefore, the Edgeworth expansion only provides an approximation to the marginal density of included data points. In this spirit, it hence leaves us the choice which data points to include. The caveat is that depending on which points are included, different truncations of the Edgeworth series may be required for optimal representation of the true distribution function. The next section will hence establish a framework for determining the optimal truncation order, conditional on the chosen data vector.


\section{Divergence and asymptoticness}
\label{Sec:Div}
In the previous section, we saw that we can approximate the distribution of any set of modes. We now have to solve the remaining question of how to judge the accuracy of a truncated Edgeworth approximation. Skillful approximations represent the approximand\footnote{Approximand: counterpart to `approximation'. The approximand is the function to be approximated. The approximation is the function that approximates the approximand.} reliably, meaning they maintain most of its important features, while not adding artefacts. Throughout physics, Taylor series are the best-known approximations and work by taking derivatives of the approximand. As most functions in physics are analytic, the Taylor series then converges towards the approximated function. In fact, analytic functions are so widespread throughout physics, that it has become conventional not to investigate the convergence of the approximation anymore: for analytic functions, an improvement in the Taylor approximation by simply adding higher-order terms is guaranteed. This has lead to a justified widespread trust that simply adding higher-order terms in approximations will improve the quality of the approximation, or at least not worsen it. This does not hold for the Edgeworth series anymore.

\begin{figure*}
\includegraphics[width=\textwidth]{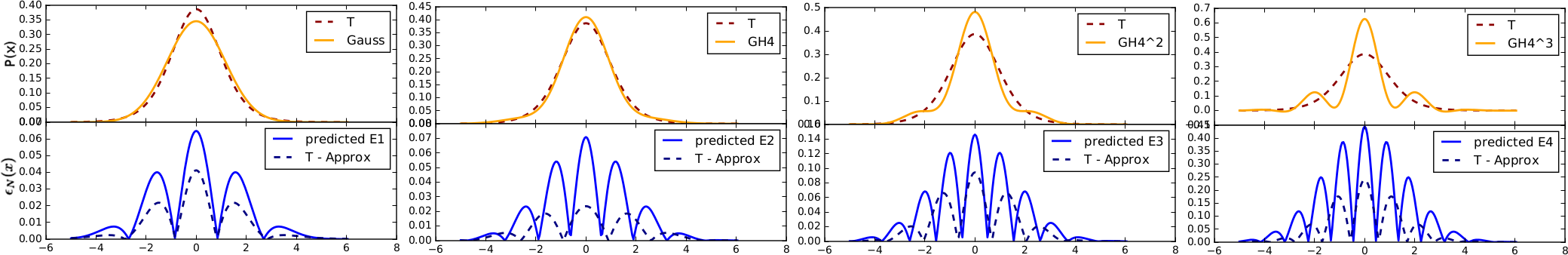}
  \caption{Example series of the local error terms $\epsilon(x)$ before being integrated over. The top-row displays a non-Gaussian target distribution (dashed), together with its Edgeworth approximations. The bottom row depicts the true (but in reality incomputable) local error $\epsilon(x)$ (dashed) and the predicted (in reality computable) errors $\hat{\epsilon}_n(x)$. The bottom row illustrates that the predicted error follows the incomputable true error remarkably well. This is to be expected from the asymptotic ordering of the Edgeworth series. The top row further illustrates the typical behaviour of the Edgeworth expansion of first improving the approximation, but then worsening it when including too many terms. In the depicted case, truncating the expansion after inclusion of $\kappa_4 H_4(x)$ yields the optimal approximation.}
\label{Fig:Texample}
\end{figure*}

The Edgeworth series is a non-convergent series. This statement is true irrespective of whether the Edgeworth series uses a Gaussian likelihood as initial approximation, or any other. For a non-convergent series, adding higher-order terms will then not automatically lead to an improvement of the approximation. Rather, adding higher-order terms can lead to a rapid deterioration of the approximation, meaning the approximation becomes ever less representative of the approximand. In order to use the Edgeworth expansion as a representation of the non-Gaussian distribution of cosmological matter fields, we therefore require a precise prediction of the approximation's error.

\begin{figure*}
\includegraphics[width=0.96\textwidth]{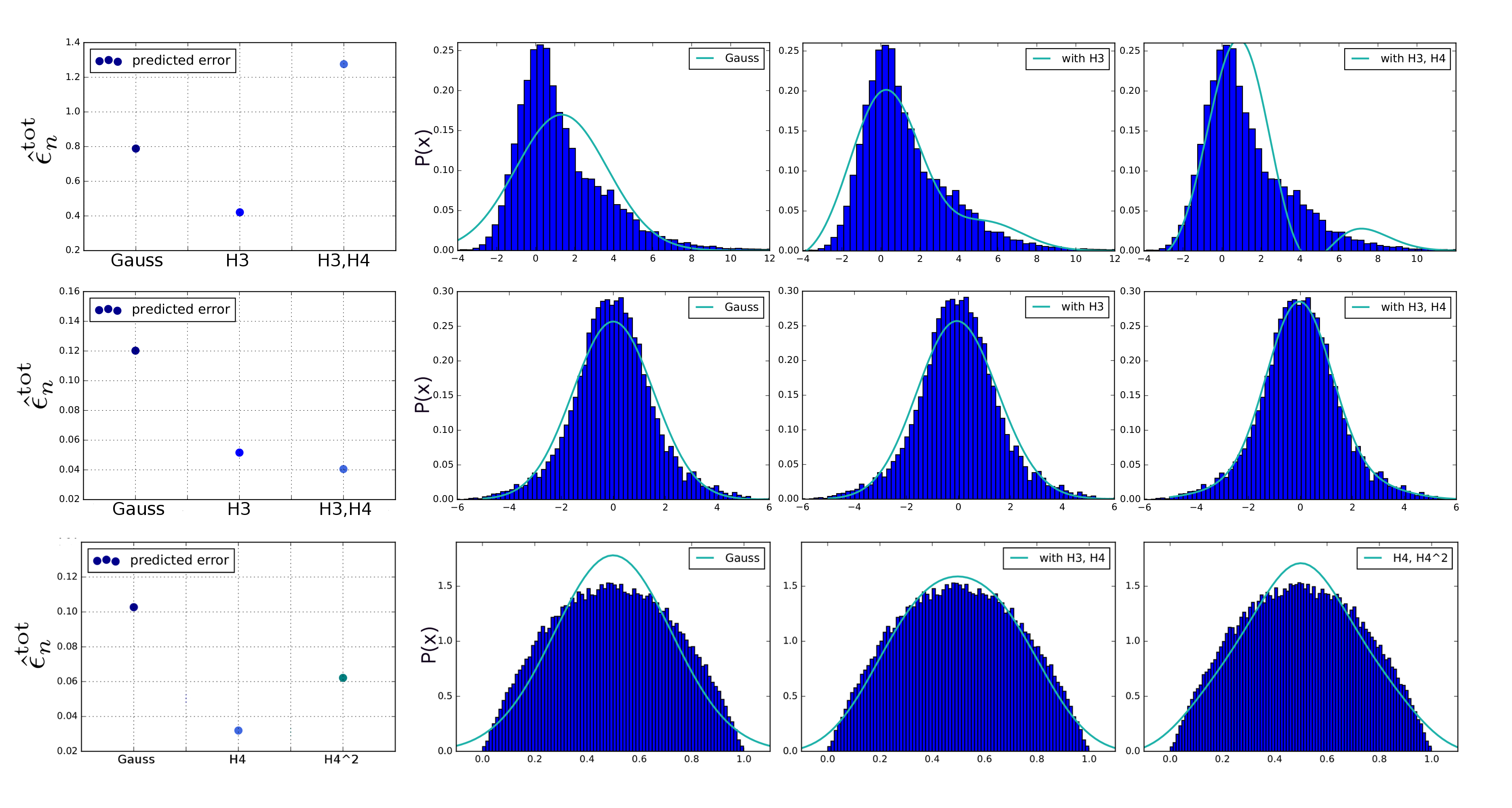}
\caption{Finding the optimal truncation point for the Edgeworth expansion, using an asymptotic error prediction. Blue depicts the target distribution, which is in reality unknown and only shown for reference here. The solid green line is the Edgeworth expansion to different orders. The aim is to find the optimal truncation point of the Edgeworth series. The first panel shows the predicted error of each order. These errors were predicted without knowing the true distribution. As can be seen, the smallest predicted errors indeed pick out the best term to truncate the Edgeworth expansion. This demonstrates that our error prediction for the Edgeworth expansion allows for a controlled expansion of unknown probability density functions. As the probability density function of cosmological matter fields is an object of physical interest, possessing its most faithful representation by cumulants is obviously of great value.}
\label{Fig:3Examples}
\end{figure*}

\begin{figure*}
\includegraphics[width=0.96\textwidth]{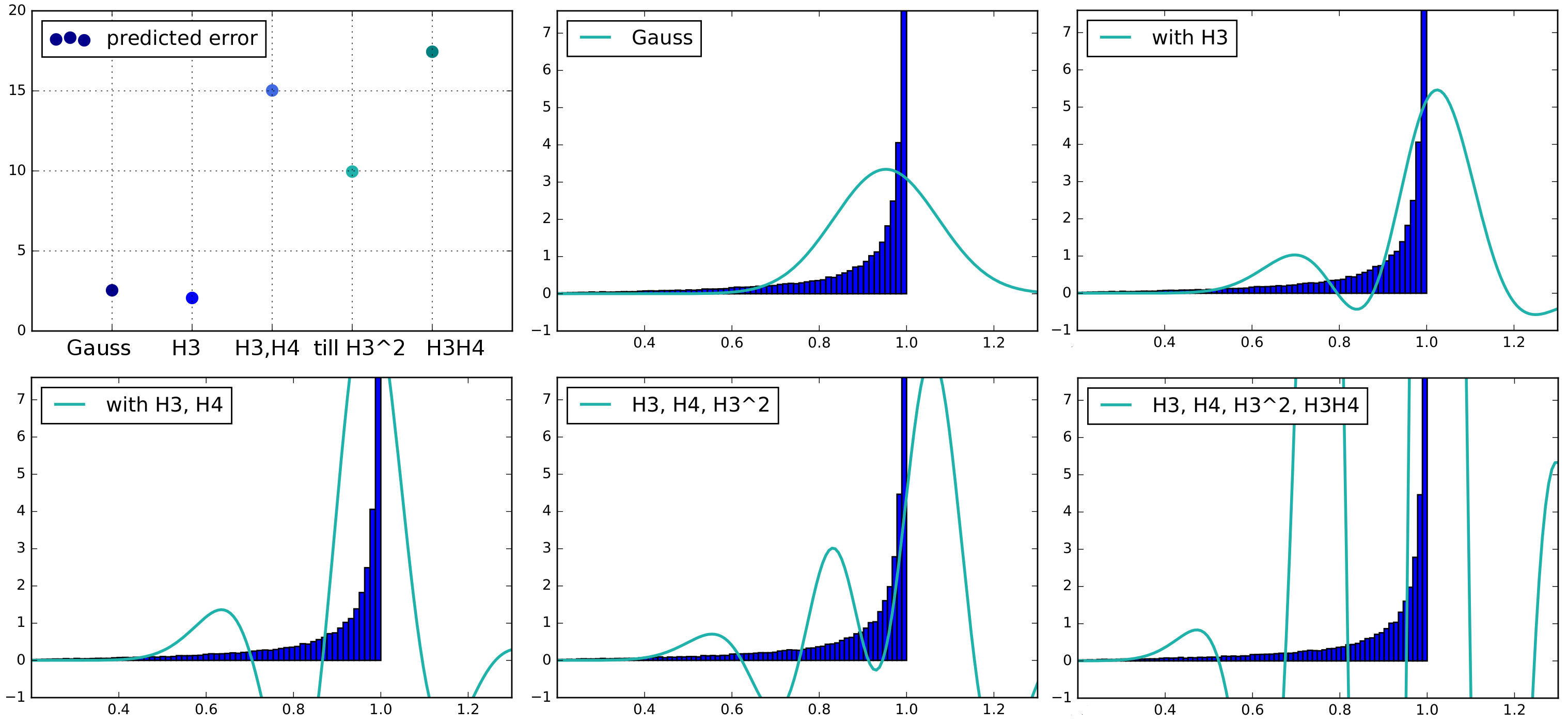}
\caption{Extreme case of the asymptotic error prediction, as Fig.~\ref{Fig:3Examples}, only now expanded to even higher-order terms. Again, the predicted errors faithfully recognize at which term the Edgeworth expansion should be truncated to yield the best representation of the unknown probability density function. The total error is then not necessarily small on an absolute base, but it is still the smallest, given the known cumulants and the Edgeworth expansion.}
\label{Fig:ExtremeCase}
\end{figure*}

For the current introduction, we loosely define the term `error' as the total difference between the approximand and the approximation. This will be made mathematically more precise in a later section, but it is already clear that a conventional error prediction requires knowledge of the approximand. In our setup, the approximand is the true non-Gaussian distribution of cosmic matter fields. This distribution is, however, unknown, thereby leaving us in the position of wanting to construct an approximation to an unknown function, and at the same time wanting to know the approximation's error with respect to the unknown function. Fulfilling these requirements is far from easy, but remarkably enough, possible. Due to the mathematical novelty, we will begin with a textual and illustrative description, which is then followed by the mathematical derivations. Further mathematical background is given in Appendices \ref{App:EW} and \ref{App:Asymptotic}.

The Edgeworth expansion can be derived from a term-by-term inversion of an integral transform of the cumulant generating function; see Appendix \ref{App:EW}. Such a term-by-term inversion of integral transforms often gives rise to a special subbranch of divergent series, called \emph{asymptotic} series. Such asymptotic series sum up functions $f_n(x)$, meaning each term of the series is again a function of a variable, and the entire series then sums those functions
\begin{equation}
 \sum_{n = 0}^\infty f_n(x).
\end{equation}
The decisive property of asymptotic series is that the sequence of functions $f_n(x)$ obeys a special ordering. The ordering thereby results in the error of the approximation being bounded to the same order of magnitude of the first omitted function. Importantly, the error of the asymptotic approximation of a function is typically a \emph{non}-monotonic function of the order $n$ up to which terms are included. For a divergent asymptotic series, including only the first $n_i$ terms typically decreases the error. However, if more than $n_i$ terms are added, then the error typically increases and often extremely rapidly so. Consequently, there exists an optimal number of terms $n_{\rm opt}$ to be included, which will yield the smallest discrepancy between approximation and approximand. Finding this optimal truncation point is therefore of great interest, and we will in the following demonstrate how this can be achieved.

Concerning the Edgeworth expansion, it consists of 3 ingredients: firstly, the Gaussian distribution, around which the expansion is made; secondly, the cumulants of the target distribution function; thirdly the Hermite polynomials in the univariate case, or Hermite tensors in the multivariate case.

Specifying the cumulants to the cumulants of other distribution functions allows us to approximate a wide range of distribution functions to variable degrees of precision. Each of these approximations will differ by the cumulants used, but all of them will contain the same Hermite polynomials. It is this sequence of Hermite polynomials, and their asymptotic ordering, that we will exploit for a remarkably accurate prediction that never needs the unknown distribution of matter fields to be specified.

Fig.~\ref{Fig:3Examples} demonstrates this error prediction. We populate a histogram with random samples, drawn from a mixture of probability density functions that are unknown to us. This histogram represents an arbitrarily shaped probability density function which is to be approximated with the Edgeworth expansion. In order to mimic the situation in cosmology, where the target probability density function is unknown, we have blinded the histograms throughout the analysis. Accessible were only the sample-estimates of the variance, and the third and fourth cumulants $\kappa_3$ and $\kappa_4$.

Given the cumulants, multiple approximations based on the Edgeworth expansion are possible: the approximation can be truncated at the Gaussian level, ignoring the higher-order cumulants. At the next order, the Edgeworth expansion takes three derivatives of the zeroth-order approximation. This includes $\kappa_3 H_3(x)$. In successively increasing orders, the Edgeworth expansion afterwards includes the four-derivatives term $\kappa_4 H_4(x)$, then the six-derivatives term $\kappa_3^2 H_3(x)$, then the seven-derivatives term $\kappa_3\kappa_4 H_7(x)$ and finally the eight-derivatives term $\kappa_4^2 H_8(x)$, which is the highest order we consider. We therefore see that merely specifying the cumulants of a target distribution leaves an arbitrariness of where to truncate the Edgeworth series. This raises the question which of the possible approximations is the best representation of the blinded target distribution function.

Denoting the blinded target distribution function by $\calP(x)$ and its Edgeworth expansion to the $n$th order by $\calE_n(x)$, a suitable definition of the error of the approximation is
\begin{equation}
 \epsilon = \int |\calP(x) - \calE_n(x)| \mathd x,
 \label{err}
\end{equation}
where we abbreviate the integrand to $\epsilon(x) = |\calP(x) - \calE_n(x)|$ for later convenience. We will use Eq.~(\ref{err})  only as a benchmark of the upcoming error prediction as the error here defined can only be evaluated if the target distribution $\calP(x)$ is known, which is not the case in a cosmology-relevant setup.

Building on the theory of asymptotic series, we now define a proxy for the true error Eq.~(\ref{err}), which can be evaluated even if the target probability distribution function is unknown. By definition of asymptotic series we have that if an arbitrary function $f(x)$ is asymptotically expanded by a sequence of functions ${f_n(x)}$, then the error of the $N$th order expansion satisfies
\begin{equation}
 \epsilon_N(x) = \left| f(x) - \sum_{n=0}^N a_n f_n(x)  \right|= \mathcal{O}\left[f_{N+1}(x)\right].
 \label{epsn}
\end{equation}
Consequently the error is of the order of the first omitted function. For more mathematical detail, the reader is referred to Appendix \ref{App:Asymptotic}. In the case of the Edgeworth expansion, the Hermite polynomials or Hermite tensors take the role of the functions $f_n(x)$, and the prefactors $a_n$ are given by the cumulants. To be precise, the Edgeworth expansion is asymptotic as a function of the cumulants $\kappa_n$: for $\kappa_n \to 0 \forall n$, the Edgeworth series converges to the Gaussian distribution for all values of the random variable $x$. In contrast, the Edgeworth expansion is \emph{not} asymptotic as a function of the random variables $x$: there exists no special $x_0$ such that for $x \to x_0$ the Edgeworth expansion approximates a given non-Gaussian distribution ever more accurately.

This is very convenient, since it allows us to define proxy error terms $\hat{\epsilon}(x)$ as a function of the cumulants, instead of the entire unknown probability density. The aim of such an error proxy is to measure the discrepancy between the true probability density function and its Edgeworth approximation, for all values of the random variable $x$. The local errors $\hat{\epsilon}(x)$ can then be summed up into a total error, by integrating over the support of the random variable. By exploiting the asymptoticness of the Edgeworth series, we can ensure that the total error proxy has the same order of magnitude as the true error (which we could however not calculate). Casting this argument into mathematical terms, we wish to construct an error proxy $\hat{\epsilon}(x)$ such that
\begin{equation}
\int \hat{\epsilon}(x) \mathd x = \mathcal{O} \left(   \int |\calP(x) - \calE_n(x)| \mathd x \right).
\end{equation}

In order to achieve such an error prediction, we proceed as follows. If the Edgeworth expansion is truncated at the Gaussian level
\begin{equation}
 \calP(x) \approx \calG(x,\kappa_2)
\end{equation}
we could define as proxy for the local error
\begin{equation}
\hat{ \epsilon }(x) = \calG(x,\kappa_2) \bigg\rvert \frac{\kappa_3}{3!} H_3(x) \bigg\rvert.
\end{equation}
The absolute value is taken such that the integral Eq.(\ref{err}) monotonically increases when integrating over the sampling space $x$. This definition of an error term is in line with the properties of asymptotic series, Eq.~(\ref{epsn}), but will incorrectly be zero  for symmetric non-Gaussian distributions since these have vanishing odd cumulants. We therefore always add the first omitted even and odd terms of an expansion. This leads to the error proxy
\begin{equation}
 \hat{\epsilon}_0(x) = \calG(x,\kappa_2) \left( \bigg\rvert \frac{\kappa_3}{3!} H_3(x) \bigg\rvert + \bigg\rvert \frac{\kappa_4}{4!} H_4(x)\bigg\rvert \right).
 \label{E1}
\end{equation}
for the Gaussian approximation. For the next order expansion
\begin{equation}
 \calP(x) = \calG(x,\kappa_2)\left(1+\frac{\kappa_3}{3!}H_3(x)\right)
\end{equation}
we define the error
\begin{equation}
 \hat{\epsilon}_1(x) = \half \calG(x,\kappa_2) \left( \bigg\vert \frac{\kappa_4}{4!} H_4 \bigg\vert + \bigg\vert \frac{\kappa_3^2}{(3!)^2}H_6(x)\bigg\vert \right).
 \label{E2}
\end{equation}
For the approximation that also includes the fourth Hermite polynomial
\begin{equation}
 \calP(x)= \calG(x,\kappa_2)\left(1+\frac{\kappa_3}{3!} H_3(x) + \frac{\kappa_4}{4!} H_4(x)\right),
\end{equation}
we define
\begin{equation}
\hat{ \epsilon}_2(x) = \half \calG(x,\kappa_2) \left( \bigg\vert \frac{\kappa_3\kappa_4}{3! 4!} H_7(x) \bigg\vert + \bigg\vert \frac{\kappa_4^2}{(4!)^2} H_8(x) \bigg\vert \right).
 \label{E3}
\end{equation}
This system of assigning an error to a given approximation can be expanded to higher-orders: always the next-order even and odd terms after the truncation order of the series are added to give an asymptotically motivated proxy for the true error of the approximation. These local errors $\hat{ \epsilon}_n(x)$ are depicted in Fig.~\ref{Fig:Texample}, and as can be seen, they are closely correlated with the true discrepancy between approximation and (in reality unknown) distribution function.

The multivariate extensions of these errors are straightforward.
In the multivariate case, truncating the Edgeworth expansion at the Gaussian level leads to the asymptotic error
\begin{equation}
 \hat{\epsilon}_0(\vec{x}) = \calG(x^i, \kappa_{i,j}) \left( \bigg\rvert \frac{\kappa^{i,j,k}}{3!} \calH_{ijk}(\vec{x}) \bigg\rvert + \bigg\rvert \frac{\kappa^{i,j,k,l}}{4!} \calH_{ijkl}(\vec{x})\bigg\rvert \right).
 \label{Emv1}
\end{equation}
If the third Hermite tensor is included in the Edgeworth expansion, then the approximation reads
\begin{equation}
 \calP(x) = \calG(x^i,\kappa_{i,j})\left(1+\frac{\kappa^{i,j,k}}{3!}\calH_{ijk}(\vec{x})\right),
\end{equation}
and we define the asymptotic error to be
\begin{equation}
 \hat{\epsilon}_1(\vec{x}) = \half \calG(x^i, \kappa_{i,j})  \left( \bigg\vert \frac{\kappa^{i,j,k,l}}{4!} \calH_{ijkl} \bigg\vert + \bigg\vert \frac{\kappa^{i,j,k}\kappa^{l,m,n} [10]}{6!}\calH_{ijklmn}(\vec{x})\bigg\vert \right).
 \label{Emv2}
\end{equation}
The total error of the distribution's $n$th-order expansion throughout the sampling domain is then the integral over the random variable $x$
\begin{equation}
 \hat{\epsilon}^{\rm tot}_n = \int \hat{\epsilon}_n(x) \mathd x.
 \label{AE}
\end{equation}
This definition for a total error has a clear intuitive meaning: as $\hat{\epsilon}_n(x)$ is positive definite for all $x$, its integral will monotonically increase as a function of $x$, and thereby sum up all pointwise differences between the approximated and unknown true probability density function. Peak and tail regions of the probability density function then contribute with equal weights to the total error: independent of whether $x$ lies in the tails or in the bulk of the distribution, the local difference between the approximation and the true probability density function contributes \emph{linearly} to the total error\footnote{This error definition can quickly be generalized by introducing a weight function, if one e.g.~wishes to suppress the contributions of the tails to the total error.}. The total error will then be large, if the approximation is bad somewhere in the sampling domain. If all higher-order cumulants tend to zero, then for each point $x$, will the Edgeworth expansion tend to the Gaussian distribution. All local errors $\hat{\epsilon}_n(x)$ will then decrease, and so will the total error Eq.~(\ref{AE}).

\begin{figure*}
\includegraphics[width=0.45\textwidth]{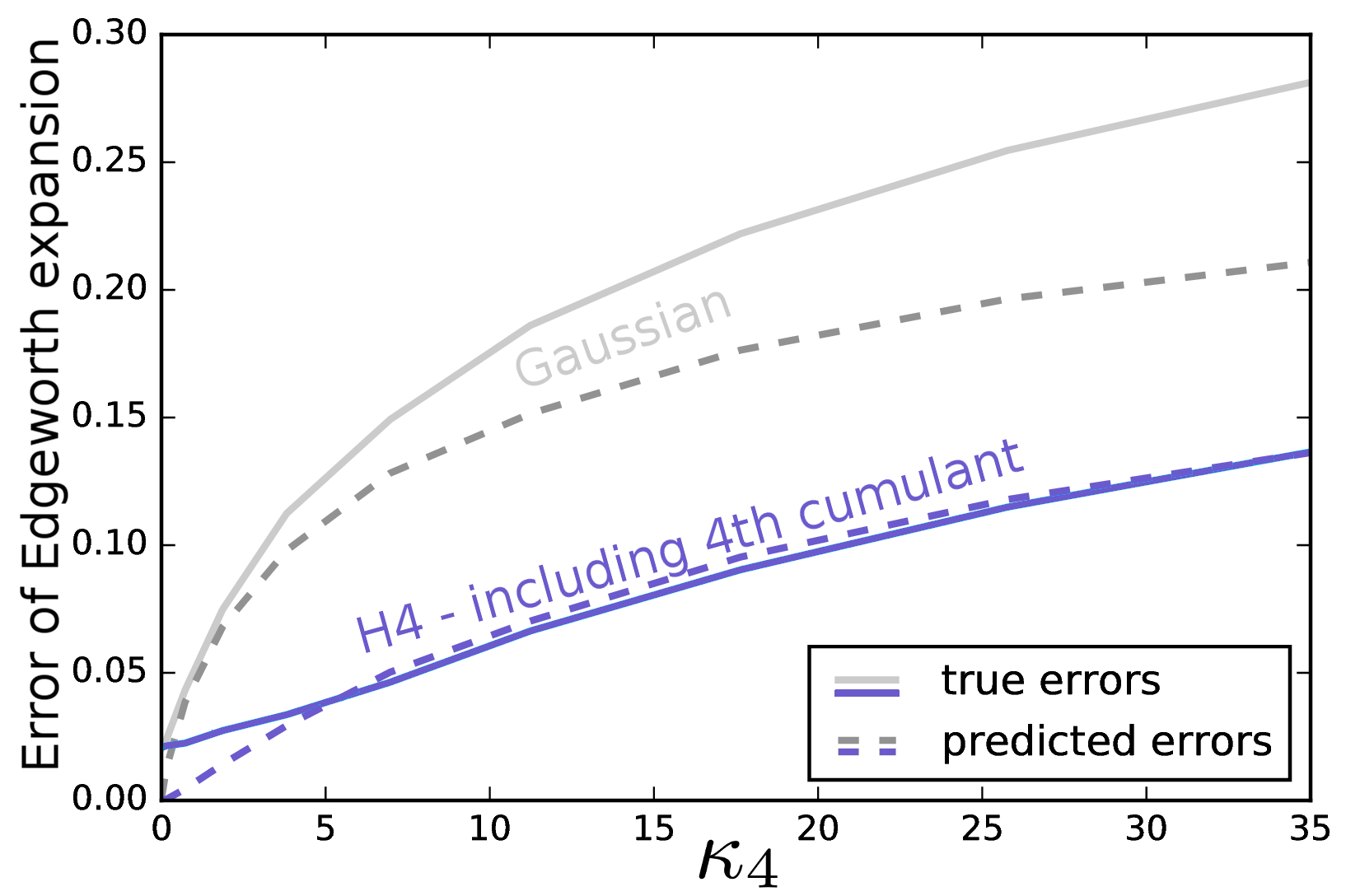}
\caption{Reliability of the error prediction for a symmetric distribution ($\kappa_3 = 0$), for increasing $\kappa_4$, always rescaled to unit variance $\kappa_2 = 1$. The predicted errors Eq.~(\ref{AE}) are plotted with dashed lines and the true errors Eq.~(\ref{err}) with solid lines. Grey refers to the errors of the Gaussian approximation, purple to the errors of the Edgeworth-term, including $\kappa_4$. Clearly visible is the tight correlation between the blindly predicted error, and the true error as calculated after unblinding the true distribution.}
\label{Fig:Error}
\end{figure*}

The important feature of Eq.~(\ref{AE}) is, that it is a proxy of the true error Eq.~(\ref{err}), but whereas the true error Eq.~(\ref{err}) can only be evaluated if the true probability density were known, the proxy Eq.~(\ref{AE}) can be evaluated without ever knowing the target non-Gaussian distribution function. In an application to cosmology this means we are now able to predict how close an Edgeworth-expanded probability distribution comes to the unknown probability distribution of cosmic matter fields.

In Fig.~\ref{Fig:Error} we demonstrate that our error proxies are extremely powerful and reliable, as the predicted errors are tightly correlated with the total difference between an unblinded true probability distribution and its approximation. We generated a wide range of non-Gaussian probability density distribution functions, either by mixing together samples from different distributions, or by rejection sampling. These samples were histogrammed, and we computed the cumulants of these histograms. From the cumulants, we computed the error proxy Eq.~(\ref{AE}) for different orders of the Edgeworth expansion, as indicated by the line style in Fig.~\ref{Fig:Error}. These predicted errors are indicated with dashed lines. They trace the actual error extremely well, which is plotted in solid lines, and which was calculated by integrating the difference between the histograms and the Edgeworth expansion.

Examples of our test histograms can be seen in Figs.~\ref{Fig:Texample} and \ref{Fig:3Examples}. Our test histograms for the error prediction include cases where the Gaussian distribution was the best representation of the unknown non-Gaussian likelihood. This can happen, if the Edgeworth series diverges so quickly that already inclusion of the third cumulant leads to strong pathological behaviour. However, very often the Edgeworth expansion is capable of reliably approximating distributions of surprisingly strong non-Gaussianities, and our error prediction is able to distinguish cases where the Edgeworth expansion succeeds from where it fails. The error prediction here presented also takes care of negative probability densities: the error definition Eq.~(\ref{err}) will penalize negativive probability densities, as the distance to the correct positive-definite probability density function will then be large. In this manner, the negative probabilities are included amongst the incorrect shapes produced by a truncated Edgeworth expansion, and a small predicted error will hence also imply that negative likelihoods will occur only rarely.

Applying this error prediction to realistic cosmological setups will allow us to determine on which scales the Edgeworth expansion can be used to reliably approximate the true non-Gaussian probability distribution of the cosmic matter fields. However, to compute these proxy errors, the bi- and trispectrum of the cosmic density fields need to be accurately known. In the upcoming years, these polyspectra will be measured from numerical simulations, which has the further advantage that the simulations also provide synthetic data sets to test our methods on. We hence postpone the application to cosmology until these simulation products become available.


\section{Numerical Implementation}
A numerical implementation of the multivariate Edgeworth expansion is a highly non-trivial task, since the index notation
\begin{align}
\calP(\fatx) \approx  \calG(x^i,x^j, \kappa_{i,j})[1 & + \frac{\kappa^{i,j,k}}{3!}\calH_{ijk}
+ \frac{\kappa^{i,j,k,l}}{4!}\calH_{ijkl}
+ \frac{\kappa^{i,j,k}\kappa^{l,m,n}[10]}{6!}\calH_{ijklmn} + \cdots
\end{align}
has a notational brevity that does not translate onto short evaluation times. If the contraction over repeated indices is implemented via for-loops, a single call to the likelihood scales with $d^3, d^4$ or $d^6$ for $d$ data points if the Edgeworth series is truncated after the third, fourth or sixth order respectively.

\begin{figure*}
\includegraphics[width=0.8\textwidth]{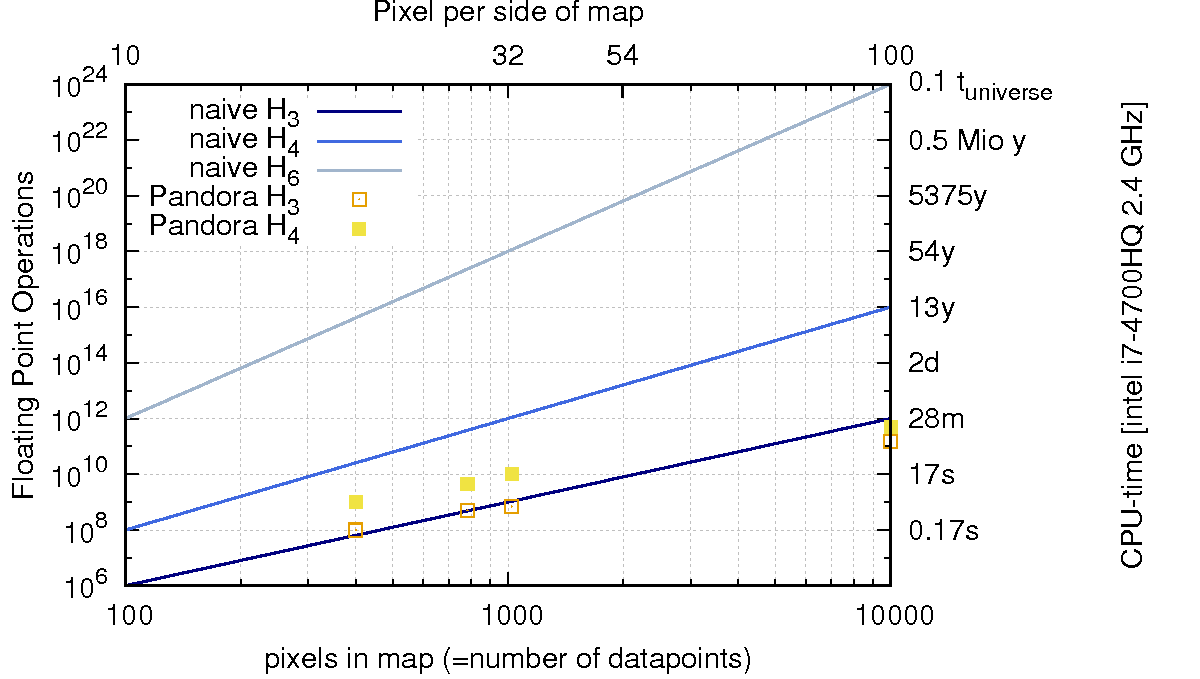}
  \caption{Numerical complexity of the multivariate Edgeworth expansion. The solid lines depict the complexity if the contraction of repeated indices is implemented as a for-loop. The open symbols depict the optimized implementation in our code `Pandora'. The right y-axis is in single-core CPU-time, ranging from seconds, over minutes, days and years to 10\% of the Universe's age (13.7 billion years). We conclude that reaching a physically interesting setup is numerically at the edge of feasible, but possible, especially if further physically-motivated simplifications to the choice of data vector could be found.}
\label{Fig:Numerics}
\end{figure*}

Parallellization of the for-loops is of course possible, but is not sufficiently effective to achieve code evaluation times below days. However, an implementation of the index contraction via for-loops leads to a large number of redundancies. Carefully resolving these leads to a massive reduction of the necessary floating-point operations, as can be seen from Fig.~\ref{Fig:Numerics}. These speed-ups were achieved in the following manner.

Given a Fourier field, our code first learns all closed polygons, for which the delta functions of a given polyspectrum are non-zero. It then replaces the sum over all indices by a sum over all solutions to the delta functions. In order for this to work, each solution has to be included with a weight that arises from distinct permutations. The code hence sorts all solutions to the delta functions with respect to the number of distinct Fourier modes per solution. This is akin to classifying by geometrical shape, e.g.~degenerate triangles, parallellograms, flat quadrilaterals, irregular quadrilaterals, etc. For each such shape, the Hermite tensors take a different form, and we provide the code with the necessary templates for the distinct forms of the Hermite tensors. Additionally, the geometrical shape of each solution to the delta function determines the number of permutations with which each solution has to be included in the sum. The code uses a lookup table to calculate a weighting prefactor, rather than summing up individual permutations. Finally, we also accounted for the Hermitian redundancy arising from the reality of the field in real space. 

For example, to compute the third Hermite tensor, which will be contracted with the bispectrum, we need to first find all triplets of Fourier modes that solve the delta function $\delta(\vec{k}_i + \vec{k}_j + \vec{k}_k)$. These can form a flat triangle, where two sides are identical, such that 3 permutations exist, or they can form a non-flat triangle, and then 6 permutations will exist. The distinct `shape templates' for the Hermite tensors begin to matter from the fourth Hermite tensor onwards. For example, given four modes that solve the delta function $\delta(\vec{k}_i + \vec{k}_j + \vec{k}_k + \vec{k}_l)$ that comes with the trispectrum, it is possible that three of the modes are identical. The closed polygon is then a flat quadrilateral, and of the fourth Hermite tensor only the term
\begin{equation}
\frac{1}{\calG} \frac{\partial}{\partial \fsub{i}} \frac{\partial}{\partial \fsub{i}}\frac{\partial}{ \partial \fsub{i}} \frac{\partial}{ \partial \fsubs{j}} \mathcal{G} = \frac{\fsubs{i}\fsubs{i}\fsubs{i}\fsub{j}}{\psub{i}^3\psub{j}} 
\end{equation}
survives. In contrast, for a paralellogramm like configuration of the polygon, the fourth Hermite tensor takes the form
\begin{equation}
\frac{1}{\calG} \frac{\partial}{\partial \fsub{i}} \frac{\partial}{\partial \fsubs{i}}\frac{\partial}{ \partial \fsub{j}} \frac{\partial}{ \partial \fsubs{j}} \mathcal{G} = \frac{\fsubs{i}\fsub{i} \fsubs{j}\fsub{j}}{\psub{i}^2\psub{j}^2} - \frac{\fsubs{j}\fsub{j}}{\psub{i}\psub{j}^2} - \frac{\fsubs{i}\fsub{i}}{\psub{i}^2\psub{j}} + \frac{1}{\psub{i}\psub{j}},
\end{equation}
and further shapes exist.

Deriving shape templates for the Hermite tensors, computing the combinatorial prefactors and taking care of the Hermitian redundancy by hand, increases the necessary coding from initially 8 lines for a summation with for-loops, to now 17,000 lines. Given the speed-ups seen in Fig.~\ref{Fig:Numerics}, we consider this effort worthwhile: a computation that would otherwise require 13 years now requires less than half an hour.

The downside of this implementation is that the code needs to keep the distinct geometrical solutions to the delta functions in memory, together with the values of the data vector, the power-, bi- and tri-spectrum. In our current implementation, the code begins from a square map in real space and uses all pixels as data points. A typical university cluster with about 100 GB RAM then allows analysis of fields with about 256 pixels per side before its RAM is exhausted. A square input map is however not necessarily the optimal dataset to start with in order to investigate non-Gaussianity: Many configurations of the bispectrum will then couple large and small scales, whereas one typically would like to split physical scales. This can for example be achieved with theory-motivated window functions such that an optimized data vector is then provided, which is adapted to scales preferred by the bi- or trispectrum. The code is able to digest such a data vector, which can reduce the necessary RAM without affecting the physical understanding. These modifications will be undertaken when we have access to realistic polyspectra of the late-time cosmic fields, as these spectra are a necessary prerequisite for this optimization.

\section{Summary}
The cosmic density fields in the late Universe are the target of current and upcoming surveys. It is often stated that cosmology endeavours to reproduce the success of the Planck mission \citep{Pl} now with these three-dimensional fields at low redshifts. However, one crucial difference between the cosmic microwave background and the late cosmic fields is that the CMB is compatible with a Gaussian field, whereas the cosmic density fields at low redshift are decidely non-Gaussian. This non-Gaussianity is brought about by non-linear gravitational collapse, and therefore is by itself already an interesting subject to study, since it improves our understanding of gravitational structure formation. Furthermore, the non-Gaussianity of the late cosmic fields also imprints itself onto the statistical properties of observables from the cosmic web. Hence, a second motivation to better study this non-Gaussianity is that it will improve the accuracy and the robustness of our data analysis methods.

Problematically though, non-Gaussianity is mathematically utterly difficult to handle. The only viable method we see in order to study it analytically within the next years, is the Edgeworth expansion. The Edgeworth expansion is however best known for its property of predicting negative probabilities, and is therefore often evaded. We have hence investigated this problem in more detail, and in fact found that it is just one of multiple problems of the Edgeworth series -- in total however, we found that the Edgeworth series arises from a mathematical framework (named asymptotic series), which is so flexible that it enabled us to a) predict when the Edgeworth expansion will fail and b) evade this situation by defining a suitable data vector. 

This implies that despite its known pathologies, the Edgeworth expansion can be used to reliably capture physical information on late cosmic fields. However, it does not do so \emph{automatically}. Rather, great care has to be taken when choosing the data vector and the truncation order. The methods described in this paper allow us to objectively assess which scales can be reliably described with this series, and which cannot. More concretely, we have established a method to predict the error that the Edgeworth series makes when approximating an unknown probability density function. We were able to predict this error blindly, meaning without knowing the non-Gaussian function. It is therefore now possible to use the Edgeworth expansion in an objectively controlled manner. 

A second result is to clarify which modes should or may be included in the Edgeworth expansion.  Since the non-Gaussianity couples all modes in principle, it is not obvious that one could apply Edgeworth to some arbitrary subset of the modes.  However, it is indeed permitted, and the Edgeworth expansion then describes the joint distribution of the included modes, with the excluded modes marginalised over. This is precisely what one would want, and Edgeworth achieves this at the same level of approximation as the expansion itself.

This paper has focused on how reliably the \emph{sampling} distribution of a data vector can be described with the Edgeworth series.
On the next level, the cumulants of the Edgeworth expansion usually depend on model parameters $\both$, and we would then like to use the expansion as a likelihood $\calP_N(\fatx|\both)$ to infer those parameters. However, in a likelihood-based inference, the likelihood itself has to be known without ambiguity, such that it takes a definite value for each parameter value. The Edgeworth expansion by itself does not yield such a unique likelihood which implies that the maximum-likelihood estimators of each truncation order will not coincide, and neither will the parameter credible regions. However, with the methods here established, the best approximation can now be singled out, whereby the likelihood can be made unique, and its remaining difference with respect to the true likelihood can now also be computed. These are necessary control elements when employing the Edgeworth series for parameter inference.  In a follow-up paper, we will study in detail how the Edgeworth series can be employed in a framework of inference.

An application of our results to a realistic cosmological setup first requires that the bi- and tri-spectrum of the evolved density fields are known more accurately. Estimating these from simulations is work in progress, as they are also required for the covariance matrices of current surveys. We will hence return to a cosmological application, when these polyspectra and their according simulated large-scale structure fields are available.

\appendix


\section{Introduction to multivariate cumulants}
\label{App:cumulants}
These appendices collect together material on multivariate statistics. Influential literature was \citet{MacKay:2003,Kendall:1977,Jeffreys:1961,Gregory:2005,McCullagh,Ghosh:2006,Box:1992}.
We are used to the concept of moments from univariate statistics. If we have $x \sim \calP(x)$, then the $m$th moment is defined as the expectation value $\left<(x)^m\right> = \int (x)^m \calP(x) \mathd x$. We will now update this to a vector-variate random variable $\fatx = (x^1, x^2, \ldots, x^n)$, where the $x^i$ are the components of $\fatx \sim \cal\calP(\fatx)$. Super-scripts denote components, whereas powers of components will either be written as multiplications, or will be denoted with brackets.

In the multivariate case, not only moments for the individual components $\left< (x^i)^m\right>$ exist, but there will be additional cross-moments $\left<x^i x^j\right>, \left<x^i x^j x^k\right>, \left<x^i x^j x^k x^l\right>$, which describe the statistical inter-dependence between the different multiplets of the components of $\fatx$. The moments of individual components encode information about the shape of that component's marginal distribution. The multivariate moments encode shape information about the joint probability distribution.

Now let there be such a probability density function $\calP(\fatx)$. Denote the average $\int f(\fatx) \calP(\fatx)\mathd^n x$ as $\left<f(\fatx)\right>$. Then we define the multivariate non-central moments as
 \begin{align}
  \kappa^i &= \left<x^i\right>\\
  \kappa^{ij} &= \left<x^i x^j\right>\\
  \kappa^{ijk} &= \left<x^i x^j x^k\right>.
 \end{align}
The covariance matrix is then given by
\begin{equation}
 \kappa^{i,j} = \left<x^i x^j\right> - \left<x^i\right> \left<x^j\right> = \kappa^{ij} - \kappa^i \kappa^j,
\end{equation}
and its inverse is denoted with downstairs indices, $\kappa_{i,j}$. The use of commas to separate indices indicates a cumulant, instead of a moment. The distinction will be made precise below.
We shall use Einstein's summation convention over repeated upper and lower indices.

A tensor-variate polynomial can then be expressed as
\begin{equation}
 a_ix^i + a_{ij}x^i x^j + a_{ijk} x^i x^j x^k + \cdots
\end{equation}
such that the $a_i, a_{ij}, a_{ijk}, \ldots $ are tensorial prefactors which are then contracted with the variables $x^i$.
We now define the multivariate moment-generating function
\begin{equation}
 M_x(\xi) = 1 + \xi_i \kappa^i + \frac{\xi_i \xi_j \kappa^{ij}}{2!} + \frac{\xi_i \xi_j \xi_k \kappa^{ijk}}{3!} + \cdots
\end{equation}
If we assume that this series converges at least for sufficiently small $|\xi|$, then we can identify the power series as the exponential series and write
\begin{equation}
 M_x(\xi) = \left< \exp(\xi_i x^i) \right>,
 \label{mom}
\end{equation}
where the expectation value transforms powers $x^i x^j\cdots x^n$ into moments $\kappa^{ij\cdots n}$. These moments are the partial derivatives $\partial_i\partial_j\cdots\partial_n$ of the moment-generating function at the origin $\xi = 0$.
Explicitly writing out the expectation operator in Eq.~(\ref{mom}), leads to
\begin{equation}
 M_x(\xi) = \int \exp(\xi_i x^i) \calP(\fatx) \mathd^n x,
\end{equation}
meaning the moment-generating function is the Laplace transform of the probability density function.

The \textbf{cumulant-generating} function is then the logarithm of the moment-generating function
\begin{equation}
 K_x(\xi) = \log M_x(\xi).
\end{equation}
Expanding the logarithm produces the series expression
\begin{equation}
 K_x(\xi) = \xi_i \kappa^i + \frac{\xi_i \xi_j \kappa^{i,j}}{2!} + \frac{\xi_i \xi_j \xi_k \kappa^{i,j,k}}{3!} + \cdots,
\end{equation}
where the cumulants are $\kappa^{i,j}, \kappa^{i,j,k}\cdots\kappa^{i,j,\ldots ,n}$ have commas between their indices while the moments do not. For real-valued $\xi$, both the moment-generating and the cumulant-generating function may be divergent. Even if all moments or cumulants exist and are finite, this does not automatically imply that the probability density function of $\fatx$ can be reconstructed. Multiple issues can arise:
\begin{enumerate}
 \item The moment- or cumulant-generating function may not be analytic at the origin. In this case, multiple probability density function can give rise to the exact same infinite set of moments. In order for an infinite set of moments to originate from a unique density function, the moment- or cumulant generating function must be analytic at the origin.
 \item The moment- or cumulant-generating function may be a non-convergent series.
 \item The inverse transform from the moment- or cumulant-generating function back to the probability density function may not possess a closed-form expression. This means a vast class of probability density functions exist which have a unique mapping to moments, but no analytical closed-form expression.
  \item If the moment- or cumulant-generating function are replaced by their truncated series expressions, the inverse transform back to the probability density function is usually not a convergent integral. This issue will be intensely investigated in this paper.
\end{enumerate}

In this index-notation, the multivariate Gaussian distribution is given by
\begin{equation}
\calG(x^i,x^j, \kappa^i, \kappa^j, \kappa_{i,j}) =  \frac{1}{\sqrt{(2\pi)^p|\kappa^{i,j}|      }} \exp\left[ -\half (x^i - \kappa^i)\kappa_{i,j}(x^j - \kappa^j) \right].
\end{equation}
If the random variable is mean-subtracted, we will write $\calG(x^i,x^j, \kappa_{i,j})$.

Cumulants and moments satisfy recursion relations with respect to each other, meaning the full sets of all cumulants and moments carry intrinsically the same information on a distribution's shape. However, many calculations with cumulants are shorter. This can be readily understood from the Gaussian distribution, whose cumulant-generating function terminates after the first two non-zero cumulants (its mean and its covariance matrix), but its moment-generating function is an infinite series due to the infinite number of non-zero even moments. In general, a distribution has either precisely two non-vanishing cumulants (then it is the Gaussian), or it has infinitely many non-vanishing cumulants.

The recursion relations between the \emph{non}-central moments and the cumulants at the lowest order are given by
\begin{align}
 \kappa^{ij} & = \kappa^{i,j} + \kappa^i \kappa^j\\
 \kappa^{ijk} & = \kappa^{i,j,k} + \kappa^i \kappa^{j,k}[3] + \kappa^i \kappa^j \kappa^k\\
 \kappa^{ijkl} & = \kappa^{i,j,k,l} + \kappa^i \kappa^{j,k,l}[4] + \kappa^{i,j}\kappa^{k,l}[3] + \kappa^i \kappa^j \kappa^{k,l}[6] + \kappa^i \kappa^j \kappa^k \kappa^l.
 \end{align}
The square brackets denote distinct permutations, for example
\begin{equation}
 \kappa^{i,j}\kappa^{k,l}[3] = \kappa^{i,j}\kappa^{k,l} + \kappa^{i,k}\kappa^{j,l} + \kappa^{i,l}\kappa^{j,k}.
\end{equation}
Up to third order, the cumulants are therefore identical to the \emph{central} moments, but differ from order 4 onwards. In the inverse direction, we have
 \begin{align}
  \kappa^{i,j} & = \kappa^{ij} - \kappa^i \kappa^j \\
  \kappa^{i,j,k} & = \kappa^{ijk} - \kappa^i \kappa^{jk} [3] + 2 \kappa^i \kappa^j \kappa^k\\
  \kappa^{i,j,k,l} & = \kappa^{ijkl} - \kappa^i \kappa^{jkl}[4] - \kappa^{ij}\kappa^{kl}[3] + 2 \kappa^i \kappa^j \kappa^{kl} [6] -6\kappa^i \kappa^j \kappa^k \kappa^l,
 \end{align}
where the moments are again non-central moments. For a symmetric distribution, all odd cumulants will vanish. Yet, if all odd cumulants vanish, the distribution can still be asymmetric, if the moment- or cumulant-generating functions are not analytic at the origin. Examples of asymmetric distributions whose odd cumulants all vanish are known. The cumulants of a sum of independent random variables are the sum of the cumulants, and if two random variables are independent, all their cross-cumulants are zero.


\section{Characteristic function}
\label{App:EW}
Above we had seen that convergence of the integrals for the moment- and cumulant-generating functions can be an issue. Convergence can be improved by changing from real-valued $\xi$ to complex $\xi$. The moment-generating function is then replaced by the \textbf{characteristic function}.  We extend the domain of the moment-generating function to the complex plane and define the characteristic function
\begin{equation}
  C_x(\xi) = \int \exp\left(\ri\xi_i x^i\right) \calP(\fatx) \mathd^n x,
\end{equation}
meaning the characteristic function is the forward (+) Fourier transform of the probability density function.
The moments are then given by
\begin{equation}
\partial_i \partial_j \cdots\partial_n C_x(\xi)\vert_0 = \ri^n \left<x^i x^j\cdots x^n\right>
\end{equation}
Formally, the probability density function can then be regained by an inverse Fourier transform:
\begin{equation}
 \calP(\fatx) = \frac{1}{(2\pi)^d}\int \exp\left(-\ri\xi_i x^i\right) C_x(\xi) \mathd^n \xi.
\end{equation}
The emphasis here has to lie on the word `formally', since the characteristic function may not be integrable, such that this backwards Fourier transform does (in general) not lead to a convergent integral.

The complex-valued cumulant-generating function is then again the logarithm of the characteristic function
\begin{equation}
 \bar{K}_x(\xi) = \log C_x(\xi) = \log \left< \exp\left(\ri \xi_i x^i\right) \right> = \sum_{n= 1}^\infty \frac{\ri^n}{n!}  \xi_i\cdots\xi_n\kappa^{i,\ldots,n} .
\end{equation}
We can now apply a sequence of mathematical operations on the last identity, in order to get around the usually non-convergent integral of the backwards Fourier transform. After exponentiation
\begin{equation}
 \left< \exp\left(\ri \xi_i x^i\right) \right> = \exp\left( \sum_{n= 1}^\infty \frac{\ri^n}{n!}  \xi_i\cdots\xi_n\kappa^{i,\ldots ,n} \right).
\end{equation}
We now express the expectation value as the integral over the density function
\begin{equation}
 \bigintssss \exp\left(\ri \xi_i x^i\right) \calP(\fatx) \mathd^nx = \exp\left( \sum_{n= 1}^\infty \frac{\ri^n}{n!}  \xi_i\cdots\xi_n\kappa^{i,\ldots ,n} \right),
\end{equation}
and apply the inverse Fourier transform:
\begin{equation}
 \calP(\fatx) = \bigintssss \exp\left( \sum_{n= 1}^\infty \frac{\ri^n}{n!}  \xi_i\cdots\xi_n\kappa^{i,\ldots ,n} \right)  \exp\left( - \ri \xi_i x^i\right) \mathd^n \xi.
\end{equation}
This can be recast into a series expression using the exponential series expansion:
\begin{equation}
 \calP(\fatx) = \bigintss \sum_k\left[ \frac{1}{k!} \left(\sum_{n= 1}^\infty \frac{\ri^n}{n!}  \xi_i\cdots\xi_n\kappa^{i,\ldots ,n} \right)^k \right]  \exp\left( - \ri \xi_i x^i\right) \mathd^n \xi.
\end{equation}
We know that doing this integral is far from trivial, and in general it will be a non-convergent function. Therefore, we now use the usual trick that Fourier transformations can turn algebraic operators into differential operators and vice versa. We have
\begin{align}
 &\frac{\partial}{\partial x^i} \exp\left(-\ri\xi_i x^i \right) = -\ri \xi_i \exp\left(-\ri \xi_ix^i\right)\nonumber\\
 &\Rightarrow \xi_i \rightarrow \frac{1}{-\ri} \frac{\partial}{\partial x^i} = \ri \frac{\partial}{\partial x^i}.
 \end{align}
We therefore replace all $\xi_i$ in the series by differential operators $\ri \partial/\partial x^i$ and arrive at
\begin{equation}
 \calP(\fatx) 
 = \bigintsss  \exp \left(\sum_{n= 1}^\infty \frac{(-1)^n}{n!} \frac{\partial}{\partial x^i}\cdots\frac{\partial}{\partial x^n}\kappa^{i,\dots ,n} \right)  \exp\left( - \ri \xi_i x^i\right) \mathd^n \xi.
\end{equation}
Next we exploit he fact that a distribution has precisely two or infintiely many non-zero cumulants. We also know that we can Fourier-transform a Gaussian, so we single out the first 2 terms of the sum over $n$ and get
\begin{align}
 \calP(\fatx) = \bigintsss  \exp \left(\sum_{n= 3}^\infty \frac{(-1)^n}{n!} \frac{\partial}{\partial x^i}\cdots\frac{\partial}{\partial x^n}\kappa^{i,\ldots ,n} \right)   \exp\left( -\kappa^i \frac{\partial}{\partial x^i} + \half \kappa^{i,j} \frac{\partial}{\partial x^i} \frac{\partial}{\partial x^j}\right)  \exp\left( - \ri \xi_i x^i\right) \mathd^n \xi.
 \end{align}
Here, we know that $\partial /\partial x^i \rightarrow -i\xi_i$, which we shall use in the middle exponential. We also know that for a smooth function, we can pull the first exponential out of the integral because the operators $\partial/\partial x^i$ are independent of the integration over $\xi$. We then have
\begin{align}
 \calP(\fatx)  =  \exp \left(\sum_{n= 3}^\infty \frac{(-1)^n}{n!} \frac{\partial}{\partial x^i}\cdots\frac{\partial}{\partial x^n}\kappa^{i,\ldots ,n} \right) \bigintsss \exp\left( +\ri \kappa^i \xi_i - \half \kappa^{i,j} \xi_i \xi_j \right) \exp\left( - \ri \xi_i x^i\right) \mathd^n \xi.
 \end{align}
The remaining integral above is the Fourier transform of the Gaussian's characteristic function. We therefore arrive at
\begin{equation}
  \calP(\fatx)  =  \exp \left(\sum_{n= 3}^\infty \frac{(-1)^n}{n!} \frac{\partial}{\partial x^i}\cdots\frac{\partial}{\partial x^n}\kappa^{i,\ldots ,n} \right) \mathcal{G}(x^i,x^j, \kappa_{i,j})
\end{equation}
where
\begin{equation}
\calG(x^i,x^j, \kappa_{i,j})=  \frac{1}{\sqrt{(2\pi)^p|\kappa^{i,j}|      }} \exp\left( -\half (x^i - \kappa^i)\kappa_{i,j}(x^j - \kappa^j) \right).
\end{equation}

If we assume that we only have the first two non-Gaussian cumulants, $\kappa^{i,j,k}$ and $\kappa^{i,j,k,l}$, at our disposal, then the first few terms of the above series are
\begin{equation}
 \calP (\fatx) \approx\left[ 1 - \frac{\kappa^{i,j,k}}{3!} \partial_i\partial_j\partial_k + \frac{\kappa^{i,j,k,l}}{4!} \partial_i \partial_j \partial_k \partial_l + \frac{\kappa^{i,j,k}\kappa^{l,m,n}}{6!} \partial_i\partial_j\partial_k\partial_l\partial_m\partial_n [10] + \cdots\right]\mathcal{G}(x^i,x^j, \kappa_{i,j}).
 \label{OperatorVersion_App}
\end{equation}
as in Eq.~(\ref{OperatorVersion}).
From now on, we will assume that we have mean-subtracted our data set, such that $\kappa^i = 0\ \forall i$. Let us now define $h_i = x^r \kappa_{ir}$, such that $\partial_m h_i = \kappa_{m,i}$. Then we have for real valued $x^r$
 \begin{align}
 & \frac{\partial_i\calG}{\calG} = - h_i \\
 & \frac{\partial_i \partial_j \calG}{\calG} = h_i h_j - \kappa_{i,j} \\
 & \frac{\partial_i \partial_j \partial_k \calG}{\calG} = - h_i h_j h_k + h_i \kappa_{j,k}[3]\\
 & \frac{\partial_i \partial_j \partial_k \partial_l \calG}{\calG} =  h_i h_j h_k h_l - h_i h_j \kappa_{l,k} [6] + \kappa_{i,j}\kappa_{k,l} [3]\\
 & \frac{\partial_i \partial_j \partial_k \partial_l  \partial_m \partial_n \calG}{\calG} =  h_i h_j h_k h_l h_m h_n - h_i h_j h_k h_l \kappa_{m,n}[15] + h_i h_j \kappa_{k,l}\kappa_{m,n}[45] - \kappa_{i,j}\kappa_{k,l}\kappa_{m,n} [15]
 \end{align}
The first terms of the multivariate Edgeworth expansion are then
\begin{align}
\calP(\fatx) \approx \calG(x^i,x^j, \kappa_{i,j})[1 & + \frac{\kappa^{i,j,k}}{3!}(h_i h_j h_k - h_i \kappa_{j,k}[3]) \nonumber\\
 & + \frac{\kappa^{i,j,k,l}}{4!}(h_i h_j h_k h_l - h_i h_j \kappa_{l,k} [6] + \kappa_{i,j}\kappa_{k,l} [3])\nonumber\\
 & + \frac{\kappa^{i,j,k}\kappa^{l,m,n}[10]}{6!}(h_i h_j h_k h_l h_m h_n - h_i h_j h_k h_l \kappa_{m,n}[15] + h_i h_j \kappa_{k,l}\kappa_{m,n}[45] - \kappa_{i,j}\kappa_{k,l}\kappa_{m,n} [15]) + \cdots
 \label{edge}
 \end{align}
As the derivatives of the Gaussian likelihood are related to the Hermite tensors from Sect.~\ref{App:Hermi}, the Edgeworth expansion can be rewritten as
\begin{equation}
\calP(\fatx) \approx  \calG(x^i,x^j, \kappa_{i,j})\left[1 + \frac{\kappa^{i,j,k}}{3!}\calH_{ijk}
+ \frac{\kappa^{i,j,k,l}}{4!}\calH_{ijkl}
+ \frac{\kappa^{i,j,k}\kappa^{l,m,n}[10]}{6!}\calH_{ijklmn} + \cdots \right]
 \label{hermiedge_App}
\end{equation}
The last term in particular is worth considering in more detail: In the univariate Edgeworth expansion, this term would include the third cumulant squared $\kappa_3^2 H_6(x)$. Now in the multivariate case, the third cumulant is not simply squared anymore. Rather, all combinatorial possibilities of distributing six indices onto two third-order cumulants are considered. This gives rise to the 10 permutations indicated by square brackets.


\section{Introduction to asymptotic series}
\label{App:Asymptotic}
Central to the theory of asymptotic series is the concept of a \emph{function series}. A function series is a series whose summands are functions of a variable. We write $(f_n), n \in \mathbb{N}$, for the function series, and the $f_n$ are themselves functions. The brackets indicate that the entire sequence of functions is meant, whereas $f_n$ refers to the $n$th function. An example is $f_n = x^n$, such that the function series superimposes monomials. We often find the notation
\begin{equation}
 f(x) \sim \sum_{n} f_n,
\end{equation}
where the tilde `$\sim$' means the function series represents the continuous function $f(x)$ in a manner to be specified further. The function series $(f_n), n\in \mathbb{N}$ may converge pointwise or uniformly against the function $f(x)$ which it represents. The series $(f_n)$ converges pointwise against $f(x)$ if we have that
\begin{equation}
 \forall x\ \exists\  n_0  \ {\rm \ such \ that\ } \ \ \forall n\geq n_0:  |f_n(x) - f(x)| \leq \epsilon \ \ \forall \epsilon > 0 .
\end{equation}
The function series then gets arbitrarily close to the function which it represents, even though different orders $n_0$ might be needed at different points $x$ to reach the same precision. Uniform convergence is given if $n_0$ is independent of $x$. For a convergent series (in both senses), adding higher terms $f_n, n > n_0$ can only leave the remaining distance $\epsilon$ either constant, or decrease it. The precision with which the convergent series represents the original function will then automatically improve or remain constant if higher order terms are added.

For asymptotic series, the last point that is not mandatory anymore: Adding higher orders can lead to a deterioration of the the original function's asymptotic representation. If the asymptotic series is divergent, there will exist an optimal truncation which yields a minimal error.

\textbf{Small $o$ and big $\mathcal{O}$ notation:} Physics often uses the big-$\mathcal{O}$ notation which translates to `\emph{same} order of magnitude'. Similarly, the concept of $f=o(h)$ notation exists, which means $f$ is `inferior to the order of magnitude of $h$'. This can be made mathematically precise. Imagine two functions $f,h$ with $h \neq 0$. Then for $c > 0$ being a finite constant
\begin{align}
 f(x) & = \mathcal{O}(h(x)) {\rm \ for \ } x\rightarrow x_0{\rm \ if\ } \frac{f(x)}{h(x)} \leq c {\rm \ for \ } x\rightarrow x_0,\\
 f(x) & = o(h(x)) {\rm \ for \ } x\rightarrow x_0{\rm \ if\ } c \frac{f(x)}{h(x)} \rightarrow 0 {\rm \ for \ } x\rightarrow x_0.
 \end{align}
Big-$\mathcal{O}$ notation therefore indicates that the ratio of the two functions is bounded, and small $o$ notation indicates that the magnitude of the function in brackets sets an upper limit on the magnitude of the preceding function. A more rigorous definition exists which also includes the case of $h$ having a root. For many well-behaved functions, the constant $c \neq 0$ can often be worked out rather well, such that the order-of-magnitude assertion is more quantitative.

\textbf{Asymptotic series:} Let $(f_n), n \in \mathbb{N}$ be a sequence of functions, and let $f(x)$ be a continuous function.
The sequence $(f_n)$ is said to be an asymptotic sequence if the functions satisfy the relations
\begin{equation}
 \forall n\ f_{n+1} = o(f_n).
\end{equation}
The representation of $f(x)$ by the sequence $(f_n)$
\begin{equation}
 f(x) \sim \sum_{n = 0}^\infty a_n f_n(x) {\rm \ for \ } x\rightarrow x_0,
\end{equation}
is said to be \emph{asymptotic} for $x\to x_0$ if for any fixed $N$
\begin{equation}
 \epsilon_N = \left|f(x) - \sum_{n=0}^N a_n f_n(x)\right| = o(f_N(x)) {\rm \ for \ }x \rightarrow x_0.
\end{equation}
Since the sequence internally also satisfies $f_{n+1} = o(f_n)$, the error $\epsilon_N$ can also be written as
\begin{equation}
 \epsilon_N = \left|f(x) - \sum_{n=0}^N a_n f_n(x)\right| = \mathcal{O}(f_{N+1}(x)) {\rm \ for \ }x \rightarrow x_0.
\end{equation}
The error of the asymptotic representation is then inferior to the last included function, which is equivalent to it being of the same order of magnitude as the first omitted function. For asymptotic \emph{power} series this is often rewritten as
\begin{equation}
 \lim_{x \to 0} \frac{1}{x^N} \left[ f(x) - \sum_{k = 0}^N a_k x^k \right] = 0,
\end{equation}
for all $N$. This is obviously true, since by definition of asymptoticness $\left|f(x) - \sum_{k = 0}^N a_k x^k\right| \ll x^N$.

The definition of asymptoticness does not imply the error $\epsilon_N$ to be a monotonically decreasing function of $N$. In fact, asymptotic series do not only include convergent series, but also many cases of divergent series. A divergent asymptotic series often shows the behaviour of the error first decreasing with $N$, but then increasing again as the series begins to diverge. Asymptoticness is independent of convergence since convergence is a statement on $N \to \infty$, whereas asymptoticness studies the limit $x \to x_0$. Let
\begin{equation}
 S_N = \sum_{n=0}^N a_n f_n(x)
\end{equation}
be the partial sums of a series. Then convergence demands
\begin{equation}
 f(x) - S_N(x) \to 0 {\rm \ for \ } N \to \infty,
\end{equation}
meaning convergence quantifies the decreasing error for the partial sums $S_N$ as $N \to \infty$, for fixed $x$. See also the discussion of pointwise and uniform convergence above. Asymptoticness instead quantifies the boundedness of the error of $S_N(x)$, for fixed $N$ and $x \to x_0$. Often, for $x \to x_0$, the error can be made arbitrarily small, even if $N \to \infty$ can not decrease the error.

Convergent series are furthermore a series representation of a unique function. For divergent asymptotic series, this is not necessarily true anymore. Multiple distinct functions can lead to the same asymptotic series, and one function can also possess multiple distinct expansions by asymptotic series.

For $N\to \infty$, both the Edgeworth- and the Gram-Charlier series are rapidly divergent series for essentially all approximated probability distributions. The Gram-Charlier series however reorders the sequence of functions $f_n$ as given by the Edgeworth series, and thereby destroys the asymptotic ordering $f_{n+1} = o(f_n)\ \forall n$, which is still fullfilled by the Edgeworth expansion.

\section{Hermite tensors}
\label{App:Hermi}
The first 6 Hermite tensors used in this work are given as follows, using the definition $h_i = x^j\kappa_{ij}$.
 \begin{align}
&  \calH_i = h_i\\
&  \calH_{ij} = h_i h_j -\kappa_{ij}\\
&  \calH_{ijk} = h_i h_j h_k - h_i \kappa_{jk} - h_j \kappa_{ik} - h_k \kappa_{ij} = h_i h_j h_k - h_i \kappa_{jk}[3]\\
&  \calH_{ilmk} = h_m h_l h_i h_k - h_m h_k \kappa_{il}[6] + \kappa_{il}\kappa_{mk}[3] \\
&  \calH_{ijklm} = h_i h_j h_k h_l h_m - h_i h_j h_k \kappa_{lm}[10] +h_i\kappa_{jk}\kappa_{lm}[15]\\
&  \calH_{ijklmn} = h_i h_j h_k h_l h_m h_n - h_i h_j h_k h_l \kappa_{mn}[15]\\
  & \ \ \ \ \ \ \ \ \ \ \ \ \ + h_i h_j \kappa_{kl} \kappa_{mn} [45] -\kappa_{ij}\kappa_{kl}\kappa_{mn}[15]
  \label{Herms}
 \end{align}
The tensors are symmetric under index permutations, since the partial derivatives $\partial_i$ commute. The numbers in angular brackets denote permutations.

\section{Revealing the blinded distribution}
\label{Unb}
\begin{figure*}
\includegraphics[width=0.6\textwidth]{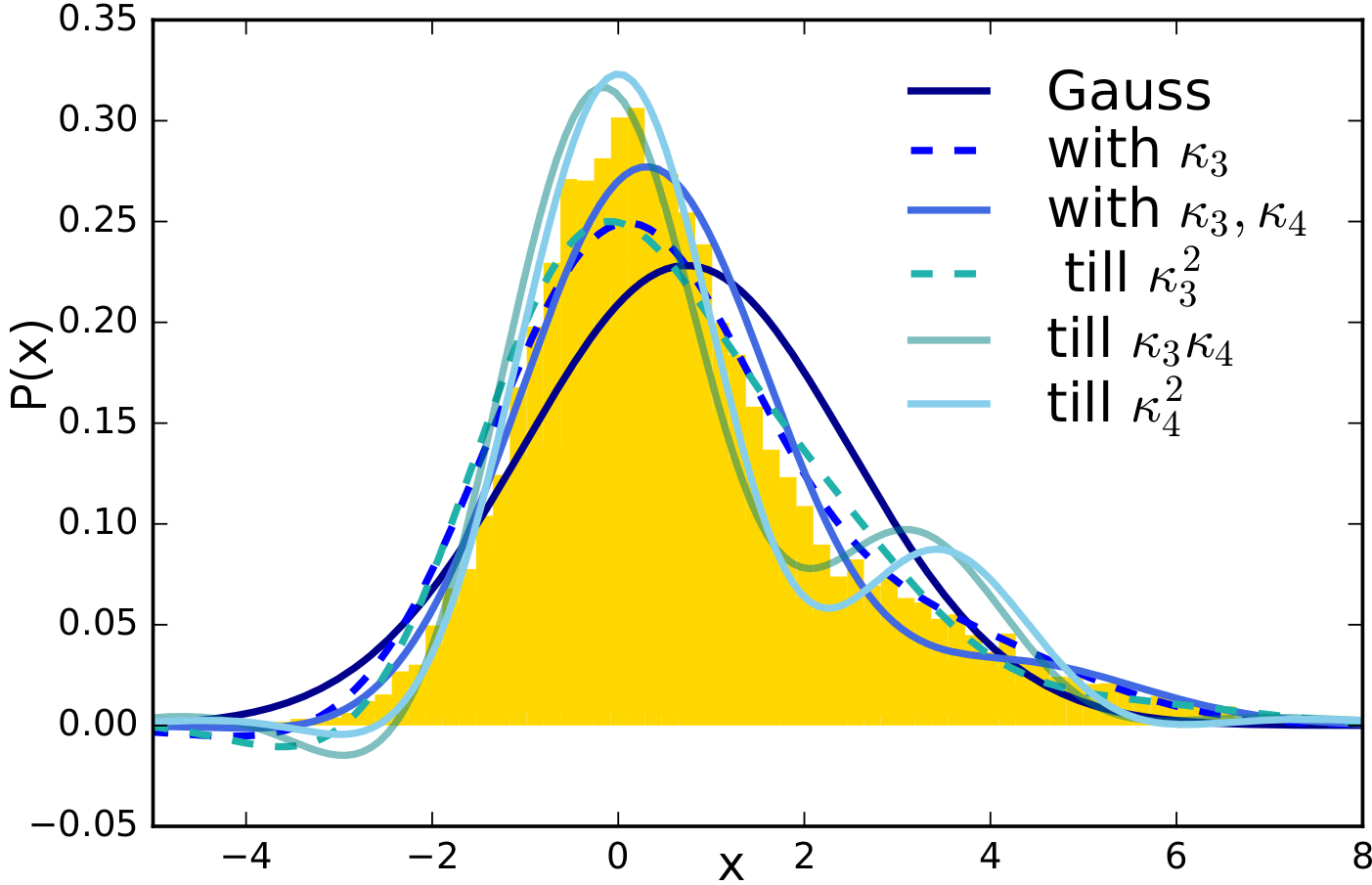}
  \caption{Revelation of the blinded probability distribution (in yellow) from Fig.~\ref{Fig:Ambiguity}. Top to bottom in the legend indicates the logical order of including higher-order terms in the Edgeworth expansion. As can be seen, stopping after the inclusion of the third cumulant times the third Hermite polynomial, $\kappa_3 H_3(x)$, would have been optimal. Including also $\kappa_4 H_4(x)$ leads to a worsening of the approximation (solid medium-blue line), including additionally $\kappa_3^2 H_6(x)$ then again improves the approximation. Afterwards, including $\kappa_3 \kappa_4 H_7(x)$ and $\kappa_4^2 H_8(x)$ introduces a fake secondary maximum and negative likelihoods to the left. The oscillation between improvement and worsening of the approximation is typical for asymptotic series, which do not have monotonically decreasing errors as a function of the included higher-orders. A method to predict the optimal truncation point is derived in Sect.~\ref{Sec:Div}.}
\label{Fig:Unblind}
\end{figure*}

For illustrative purposes only, we here reveal in Fig.~\ref{Fig:Unblind} the probability density function for which Edgeworth approximations were calculated in Fig.~\ref{Fig:Ambiguity}. In realistic cosmological applications, the probability density function cannot be revealed since it is not known. In this toy example, we see however the generic feature of the Edgeworth expansion that it easily produces incorrect shapes for the probability density function, if higher-order terms are simply added in without a careful control of the ensuing errors.

\section{Acknowledgements}
This research was funded by a DAAD research fellowship of the German Academic Exchange Service, and was supported by Grant ST/N000838/1 from the Science and Technology Facilities Council.  We thank Ruth Durrer, Martin Kunz, Licia Verde, and Raul Jimenez for support and insightful discussions.

\bibliographystyle{mn2e}
\bibliography{TDist}

\label{lastpage}
\bsp
\end{document}